\documentclass[twocolumn,aps,prl,amsmath,amssymb,amsfonts,superscriptaddress,longbibliography]{revtex4-2}

\usepackage{graphicx}
\usepackage{dcolumn}
\usepackage{bm}
\usepackage{hyperref}
\usepackage{xcolor}
\usepackage{braket}
\usepackage{dblfloatfix}
\usepackage{amsmath}
\usepackage{capt-of}
\usepackage{graphicx}

\newcommand{\PSect}[1]{\emph{#1}---}
\definecolor{hannes}{rgb}{.89,.0,.13}

\newcommand{\Az}{A_z}
\newcommand{\Bz}{B_z}
\usepackage{listings}
\usepackage{physics}
\newcommand{\ave}[1]{\langle#1\rangle}
\newcommand{\nf}{2^{-N/2}} % normalisation factor
\definecolor{pykw}{RGB}{31,119,180}   % keywords (from/import/def/class/…)
\definecolor{pycl}{RGB}{170,51,119}   % your classes
\definecolor{pycm}{RGB}{46,139,87}    % comments
\definecolor{pystr}{RGB}{214,39,40}   % strings
\definecolor{pygray}{gray}{0.45}
\lstdefinestyle{py}{
  language=Python,
  basicstyle=\ttfamily\small,
  keywordstyle=\color{pykw}\bfseries,
  commentstyle=\color{pycm},
  stringstyle=\color{pystr},
  numbers=left, numberstyle=\tiny\color{pygray}, numbersep=6pt,
  showstringspaces=false, breaklines=true, columns=fullflexible, frame=single
}
\lstdefinestyle{pyclasses}{
  style=py,
  classoffset=1,
  morekeywords={Qcycle,Qinit,Qpivot,Qmask},
  keywordstyle=\color{pycl}\bfseries,
  classoffset=0
}

\begin{document}

\title{Generating Generalised Ground-State Ansatzes from Few-Body Examples}

\author{Matt Lourens}
\email{lourensmattj@gmail.com}
\affiliation{Department of Physics, Stellenbosch University, Stellenbosch, South Africa}

\author{Ilya Sinayskiy}
% \email{author.two@example.com}
\affiliation{School of Agriculture and Science, University of KwaZulu-Natal, Durban, South Africa}
\affiliation{National Institute for Theoretical and Computational Sciences (NITheCS), Stellenbosch, South Africa}

\author{Johannes N. Kriel}
% \email{author.two@example.com}
\affiliation{Department of Physics, Stellenbosch University, Stellenbosch, South Africa}

\author{Francesco Petruccione}
% \email{author.two@example.com}
\affiliation{Department of Physics, Stellenbosch University, Stellenbosch, South Africa}
\affiliation{National Institute for Theoretical and Computational Sciences (NITheCS), Stellenbosch, South Africa}
\affiliation{School of Data Science and Computational Thinking, Stellenbosch University, Stellenbosch, South Africa}

\date{\today}

\begin{abstract}
    We introduce a method that generates ground-state ansatzes for quantum many-body systems which are both analytically tractable and accurate over wide parameter regimes. Our approach leverages a custom symbolic language to construct tensor network states (TNS) via an evolutionary algorithm. This language provides operations that allow the generated TNS to automatically scale with system size. Consequently, we can evaluate ansatz fitness for small systems, which is computationally efficient, while favouring structures that continue to perform well with increasing system size. This ensures that the ansatz captures robust features of the ground state structure. Remarkably, we find analytically tractable ansatzes with a degree of universality, which encode correlations, capture finite-size effects, accurately predict ground-state energies, and offer a good description of critical phenomena. We demonstrate this method on the Lipkin-Meshkov-Glick model (LMG) and the quantum transverse-field Ising model (TFIM), where the same ansatz was independently generated for both. The simple structure of the ansatz allows us to obtain exact expressions for the expectation values of local observables as well as for correlation functions. In addition, it permits symmetries that are broken in the ansatz to be restored, which provides a systematic means of improving the accuracy of the ansatz.
\end{abstract}

\maketitle
% \TODO{TODO:  phase transition related to eigenvalues flipping?}\newline
% \TODO{normalization LMG}
Obtaining an exact ground-state solution for an interacting quantum many-body system is generally a very difficult, if not completely intractable, task. As a result, approaches to this problem are often based on a variational ansatz, i.e. a simplified functional form of the ground state intended to capture the latter's essential physical features. A structurally simple ansatz with few parameters allows for analytic calculations and provides qualitative insights often at the expense of quantitative accuracy. Conversely, a structurally and variationally complex ansatz requires a fully numeric approach but offers improved quantitative accuracy at the expense of qualitative insight. Balancing these qualitative and quantitative extremes is challenging. Consequently, constructing ansatzes that permit an analytic treatment while yielding accurate results over a wide range of system parameters is highly desirable.

For qualitative insights, a simple ansatz is to construct a product state with minimal variational parameters, as is typical in mean-field theory (MFT) \cite{weissLhypotheseChampMoleculaire1907,betheStatisticalTheorySuperlattices1997a,onsagerElectricMomentsMolecules1936,kikuchiTheoryCooperativePhenomena1951,oguchiTheoryAntiferromagnetismII1955}. This approach and its extensions \cite{yamamotoCorrelatedClusterMeanfield2009a,wysinCorrelatedMolecularfieldTheory2000a,wysinOnsagerReactionfieldTheory2000,duExpandedBethePeierls2003,zhuravlevMolecularfieldTheoryMethod2005a,etxebarriaGeneralizedBoundaryConditions2004,mattisMolecularfieldTheoryCorrelations1979a} function by neglecting fluctuations, and offer a low-cost procedure to obtain analytic insight into a system. However, in the vicinity of critical points these fluctuations become large. Here a different method is required, such as the renormalisation group (RG) \cite{wilsonRenormalizationGroupCritical1971,fisherRenormalizationGroupTheory1998,kadanoffStaticPhenomenaCritical1967}. 

For quantitative accuracy, a powerful class of variational ansatzes is tensor network states (TNS). A special case, matrix product states (MPS) \cite{fannesFinitelyCorrelatedStates1992,perez-garciaMatrixProductState2007a,schollwockDensitymatrixRenormalizationGroup2011}, naturally represent low-energy states of systems in 1D with local interactions \cite{hastingsAreaLawOnedimensional2007,verstraeteMatrixProductStates2006a}. Other classes of network states include projected entangled pair states \cite{verstraeteMatrixProductStates2008a,verstraeteRenormalizationAlgorithmsQuantumMany2004,murgVariationalStudyHardcore2007,murgExploringFrustratedSpin2009}, the multiscale-entanglement-renormalisation ansatz \cite{vidalClassQuantumManyBody2008c}, and tree tensor networks \cite{shiClassicalSimulationQuantum2006b,silviHomogeneousBinaryTrees2010,gersterUnconstrainedTreeTensor2014,gersterFractionalQuantumHall2017}. While TNS are broadly applicable, their parameter count typically scales linearly with system size and polynomially with bond dimension, limiting analytical tractability.

For both qualitative insights and quantitative accuracy, we introduce a method to generate tensor network states with minimal structural and variational complexity while preserving high accuracy. We leverage a domain-specific-language (DSL) -- specific syntax and rules for compactly expressing TNS via modular building blocks -- implemented as an open source Python package \cite{lourensHierarqcalGithub2024,lourensHierarchicalQuantumCircuit2023}. These blocks encode the system's size scaling, spatial homogeneity and correlations. An evolutionary algorithm exploits this DSL to construct low-energy tensor network states for a given Hamiltonian. Our approach shares several complementary threads with recent works. In variational quantum algorithms, searching over circuit structures yields ansatzes with fewer gates yet comparable performance \cite{liQuantumOptimizationNovel2020b,duQuantumCircuitArchitecture2022b,bilkisSemiagnosticAnsatzVariable2023}. Using few-site information to build scalable wavefunctions has proved effective in many-body methods such as DMET \cite{kniziaDensityMatrixEmbedding2012} and variational Monte Carlo \cite{yangScalableVariationalMonte2020}, likewise in scalable variational quantum eigensolver constructions \cite{farrellScalableCircuitsPreparing2024a,farrellQuantumSimulationsHadron2024a,gustafsonSurrogateconstructedScalablecircuitsAdaptive2025}. The Tequila framework \cite{kottmannTEQUILAPlatformRapid2021} demonstrates the power of a domain-specific language and abstract data structures in the context of quantum algorithm development.\\

\begin{figure*}[!t]
    \centering
    \includegraphics[width=\textwidth]{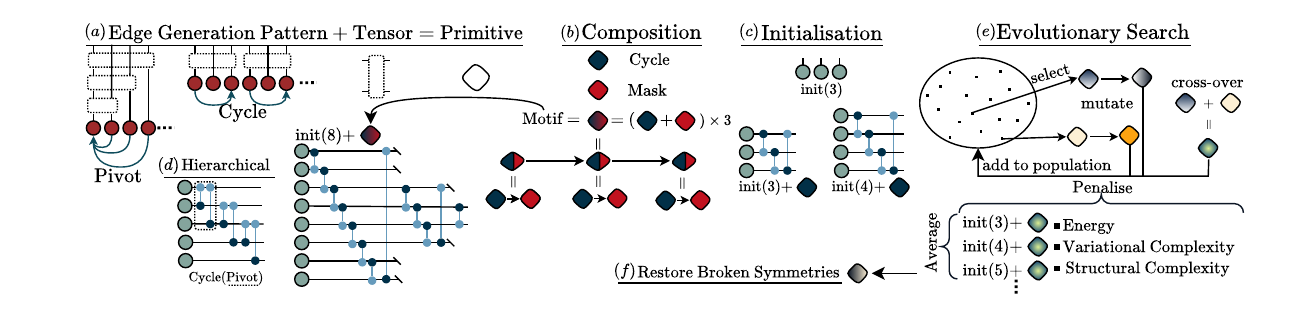}
    \caption{An overview of our method. A domain-specific-language enables ansatz generation via an evolutionary algorithm. (a) A primitive (\emph{cycle}, \emph{mask} or \emph{pivot}) is an edge generation pattern associated with a tensor. (b) Composition: Sequences of primitives form motifs; sequences of motifs form higher-level motifs. (c) A primitive applied to an initialised state (specified with the \emph{init} command) forms a tensor network. (d) A specified network, being itself a tensor, can again be associated with an edge generation pattern to form a new primitive. (e) The evolutionary algorithm mutates and crosses over motifs each generation. (f) Once the ansatz is found, broken symmetries can be restored.}
    \vspace{-20pt}
    \label{fig:gen_anz}
\end{figure*}

\PSect{Method}The elements of our method are outlined in Fig.~\ref{fig:gen_anz}. Within the DSL the elementary building blocks are called \emph{primitives}, of which we show three, a \emph{cycle, pivot} and \emph{mask}.\clearpage
\onecolumngrid
\twocolumngrid 
\noindent Examples of DSL usage is provided in Section V of the Supplementary Material (SM) \cite{lourensSupplementaryMaterial}. See also \cite{lourensHierarchicalQuantumCircuit2023,lourensHierarqcalGithub2024} for a more detailed account. In brief, the cycle and pivot primitives are defined by size-independent properties, the main two being an edge generation pattern with an associated tensor (a). Other such properties include edge order, weight sharing and boundary conditions. The mask is a special primitive which hides/unhides physical indices based on some pattern. These primitives can be composed sequentially (b) to create a sequence of primitives called a \emph{motif}. In turn, multiple motifs can be composed to form higher-level motifs, and so on. In this way motifs are built without specifying a system size, and size-scaling is captured within the properties of primitives. System size is specified through the \emph{init} command, which initialises the state with a tensor (c). When a motif is applied to an initialised state, a tensor network is formed. Since a tensor network is itself a tensor, it can again be associated with an edge generation pattern, thereby forming a new primitive, and allowing larger networks to be built hierarchically from sub-networks (d).\\ Our evolutionary algorithm (e) follows the design of those in Neural Architecture Search \cite{liuHierarchicalRepresentationsEfficient2018a,pmlr-v70-real17a} which is based on tournament selection \cite{goldbergComparativeAnalysisSelection1991}. The optimal ansatz produced by this algorithm will generally not exhibit the same symmetries as the system Hamiltonian. However, with sufficient complexity penalties the resulting structure will be simple enough for broken symmetries to be restored (f).  The algorithm starts with a randomly initialised pool of primitives. The tensors associated with these primitives are chosen from a fixed set and contain variational parameters. A tournament is then repeatedly held between a set fraction $\rho\in[0,1]$ of the total population. Here $\rho$ denotes the selection pressure, with large values facilitating exploitation. During each tournament, the winners (defined below) are used to generate offspring through mutation, cross-over and other genetic operations using the DSL. The winners are typically the two fittest individuals, but there is a probability $\epsilon\in [0,1]$ that two random individuals are declared winners instead. Large values of $\epsilon$ therefore facilitate a wider exploration of possible network structures. We evaluate the fitness  $f(\psi_a)$ of an ansatz $\psi_a$ over system sizes $n\in\mathcal{N}$ with penalties on variational and structural complexity according to
\begin{align}
    f(\psi_a)&\equiv\sum_{n\in \mathcal{N}} \bigl ( E(\psi_a^{(n)})+ l_1S(\psi_a^{(n)})+ l_2V(\psi_a^{(n)}) \bigr )w_n\label{eq:fitness}.
\end{align}
Here $E$, $S$, and $V$ are respectively the per-site energy expectation value, the structural complexity, and variational complexity of an individual. The parameters $l_1$ and $l_2$ control the penalties associated with these two complexities, while $w_n\in[0,1]$ are normalised weights assigned to the various system sizes in $\mathcal{N}$. The variational complexity $V(\psi_a^{(n)})$ of an individual is the number of variational parameters it contains. The structural complexity $S(\psi_a^{(n)})$ is the sum of the ranks of the tensors that make up the ansatz, divided by $2n$. During the search, a main worker process maintains a queue of unevaluated individuals, of which the fitnesses have not yet been calculated, together with a pool of evaluated ones. Each tournament enqueues unique individuals while duplicates only increase the multiplicities of network structures already in the pool. We define one evolutionary step as the addition of $10$ unique evaluated individuals to the pool.\\\vspace{-6pt}

\PSect{Ansatz structure and expectation values}We demonstrate our method on the Lipkin-Meshkov-Glick (LMG) model \cite{lipkinValidityManybodyApproximation1965a,meshkovValidityManybodyApproximation1965,glickValidityManybodyApproximation1965} and the quantum transverse-field Ising model (TFIM) \cite{sachdevQuantumPhaseTransitions1999}. Remarkably, the search yields the same ansatz for both models, showcasing its ability to identify network structures with some degree of universality. Only small system sizes ($\mathcal{N}=\{3,4,5\}$) were required and the search takes about $6$ CPU hours to find the ansatz. Further details on the computational cost and results for the TFIM are given in the end matter.
\begin{figure}[t]
    \centering
    \includegraphics[width=\linewidth]{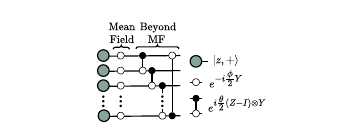}
    \vspace{-20pt}
    \caption{The ansatz generated by our method for the LMG and TFIM models.}
    \vspace{-10pt}
    \label{fig:ansatz}
\end{figure}
The ansatz is shown in Fig.~\ref{fig:ansatz}. For $N$ spins it produces the state
\begin{align}
   \ket{\psi_o}&= \left(\overleftarrow{\prod_{k=0}^{N-1}}  C^{\theta}_{k,k+1} R^{\theta}_{k+1} \right)  \left( \overleftarrow{\prod_{j=0}^{N-1}} R^\phi_j \right) \ket{z,+}^{\otimes N}, \label{eq:ansatz}
\end{align}
where $\phi$ and $\theta$ are variational parameters and $\ket{z,\pm}$ are the eigenstates of the Pauli-$Z$ matrix with eigenvalues $\pm 1$. The operator products in Eq.~\eqref{eq:ansatz} are ordered with the $k$ and $j$ indices decreasing from left to right. The two unitary operators appearing in $\ket{\psi_o}$ are
\begin{align}
	C^{\theta}_{ij} = e^{i\frac{\theta}{2} Z_i Y_j}\qquad\text{and}\qquad R^\theta_j = e^{-i\frac{\theta}{2} Y_j}.
\end{align}
Here $(X_i,Y_i,Z_i)$ are the Pauli spin matrices associated with the spin at site  $i\in\{0,\ldots,N-1\}$, obeying periodic boundary conditions: $i+N\equiv i$. As shown in Section I of the SM \cite{lourensSupplementaryMaterial}, it is possible to obtain an exact representation of $\ket{\psi_o}$ as an MPS. In this form the ansatz reads
\begin{equation}
    \ket{\psi_o}=2^{-N/2}\sum_{\vec{\sigma}} {\rm Tr}(B^{\sigma_0}A^{\sigma_1}\cdots A^{\sigma_{N-1}})\ket{y, \vec{\sigma}}, \label{eq:original_mps}
\end{equation}
where $\ket{y,\sigma_i}$ are the eigenstates of the Pauli-$Y_i$ matrix with eigenvalues $\sigma_i=\pm 1$, and
\begin{align}
    A^+ &=
    \begin{bmatrix}
        \cos{\frac{\theta}{2}} & \cos{\frac{\theta}{2}} \\[0.5em] 
        i \sin{\frac{\theta}{2}} & -i \sin{\frac{\theta}{2}}
    \end{bmatrix}
    \begin{bmatrix}
        e^{-\frac{i(\theta+\phi)}{2}} & 0 \\
        0 & e^{\frac{i(\theta+\phi)}{2}}
    \end{bmatrix}, \label{eq:A_pm}\\
    B^+ &=
    \begin{bmatrix}
        \cos{\frac{\theta}{2}} & \cos{\frac{\theta}{2}} \\[0.5em]
        i \sin{\frac{\theta}{2}} & i \sin{\frac{\theta}{2}}
    \end{bmatrix}
    \begin{bmatrix}
        e^{-\frac{i(\theta+\phi)}{2}} & 0 \\
        0 & e^{-\frac{i(\theta-\phi)}{2}}
    \end{bmatrix}, \\
    A^- &= (A^+)^*, \quad B^- = (B^+)^*.
\end{align}

Both the TFIM and LMG models exhibit translational invariance. However, this property is not shared by $\ket{\psi_o}$ due to the lone $B^{\pm}$ matrix appearing in Eq.~\eqref{eq:original_mps}. It seems that the complexity penalty on motifs during the search prevents the generation of an explicitly translationally invariant state. This coincides with the idea that symmetry-breaking ansatzes require lower structural complexity for similar ground-state energy convergence as symmetry-preserving ansatzes \cite{parkEfficientGroundState2024b, meyerExploitingSymmetryVariational2023b}. We will restore this symmetry through a minimal modification of the original ansatz $\ket{\psi_o}$ by replacing the $B^{\sigma_0}$ with $A^{\sigma_0}$ in Eq.~\eqref{eq:original_mps}. This modification marginally improves results for small systems, while still yielding the same ground-state energy as the original ansatz for both models in the thermodynamic limit.
This results in the translationally invariant ansatz 
\begin{align}
    \ket{\psi_t} &= \frac{1}{M}\sum_{\vec{\sigma}} {\rm Tr}(A^{\sigma_0}\cdots A^{\sigma_{N-1}})\ket{y,\vec{\sigma}} \label{eq:trans_mps},
\end{align}
where $M$ is a normalisation factor. The structure of both $\ket{\psi_o}$ and $\ket{\psi_t}$ in Eqs.~\eqref{eq:original_mps} and~\eqref{eq:trans_mps} allows us to obtain closed-form expressions for expectation values using a transfer matrix approach. See Section II.A of the SM \cite{lourensSupplementaryMaterial}. For $\ket{\psi_t}$ we find
%Applying this procedure to our ansatz, we obtain ($r\neq 1$):
\begin{align}
    \expval{X_i}&=\frac{1}{M^2}\left [\frac{c^2(s-t)}{st-1} +\frac{d^2(s-t)}{t^2(st-1)}(st)^N \right],\label{eq:ti_field_mag}\\
    \expval{Z_i}&=\frac{1}{M^2}\left [\frac{cd((st)^N-1) }{st-1}  \right ],\label{eq:ti_int_mag}\\
    \expval{Z_iZ_{i+r}}&=\frac{1}{M^2}\left [f(r)+(st)^Nf(-r)\right ],\label{eq:ti_interaction_corr}
\end{align}
where 
\begin{align}
	f(r)&=\frac{c^2d^2+(s-t)^2(st)^r}{(st-1)^2},\\
	M^2&=1+(st)^N, \quad |st|\neq 1
\end{align}
and
\begin{align}
	c&=\cos(\theta),\quad s=\sin(\theta),\label{eq:cs_definitions}\\
	d&=\cos(\theta+\phi),\quad t=\sin(\theta+\phi).\label{eq:dt_definitions}
\end{align}

\PSect{Results for the LMG Model}The LMG Hamiltonian for $N$ spin-$\frac{1}{2}$ particles reads
\begin{align}
    H&=-\frac{J}{4N} \sum_{i<j} Z_iZ_{j} - \frac{h}{2}\sum_{i=0}^{N-1} X_i, \label{eq:lmg_hamiltonian}
\end{align}
where $J$ and $h$ set the strengths of the spin-spin interaction and external field respectively. The all-to-all nature of the spin interaction results in the system's mean-field description becoming exact for certain predictions in the thermodynamic limit. We first show that our ansatz shares this property. Fig.~\ref{fig:ansatz} shows that our approach contains the mean-field result as a special case. Specifically, the first layer of $R^\phi$ rotations generates a product state amounting to a mean-field ansatz. When $\theta\neq0$, the second layer of $C^\theta R^\theta$ rotations then introduces correlations beyond the mean-field level. To proceed, we calculate the energy per spin in the thermodynamic limit with respect to $\ket{\psi_t}$ using Eqs.~\eqref{eq:ti_field_mag} and \eqref{eq:ti_interaction_corr}.  See Section III.A of the SM for details. This yields
%All spins interact with each other which cause the model's MFT to be exact in the thermodynamic limit. From Eq.~\eqref{eq:ansatz} the first layer of $R^\phi$ rotations creates a product state, which when minimised with respect to the Hamiltonian \eqref{eq:lmg_hamiltonian} yields the MFT solution, therefore we expect $\theta=0$  in our thermodynamic solution. From \eqref{eq:ti_field_mag} and \eqref{eq:ti_interaction_corr} the energy in the thermodynamic limit is:
\begin{align}
    \lim_{N \rightarrow \infty} \frac{\expval{H}}{N}  &=-\frac{c^2d^2}{8(st-1)^2}-\frac{h(s-t)c^2}{2(st-1)},
\end{align}
which is a function of $\theta$ and $\phi$ via Eqs.~\eqref{eq:cs_definitions} and \eqref{eq:dt_definitions}. Minimising this expression with respect to these angles produces
%\begin{equation}
	%\sin(\phi)=2h\qquad\text{and}\qquad\theta=0.
%\end{equation}
\begin{equation}
	\sin(\phi)=\begin{cases}
		2h& |2h|\leq1 \\
		{\rm sgn}(h)& \text{otherwise}
	\end{cases}\qquad\text{and}\qquad\theta=0,
\end{equation}
as shown in Section III.B of the SM. The vanishing of $\theta$ implies that our ansatz reduces to a product state generated by the first layer of $R^\phi$ rotations.
%\begin{align}
 %   s&=0\implies \theta=0 \nonumber\\
  %  h&= \frac{t}{2} = \frac{\sin \phi}{2} \label{eq:LMG_thermo_sol}
%\end{align}
Inserting this into Eq.~\eqref{eq:ti_int_mag} for $\expval{Z_i}$ yields the spontaneous magnetisation
\begin{align}
    \lim_{N\rightarrow \infty}\frac{1}{2N}\sum_i \expval{Z_i} &=\begin{cases} \pm \frac{1}{2}\sqrt{1-4h^2} & |2h|\leq1 \\
    0 & \text{otherwise}
    \end{cases}\label{eq:LMG_Magnetisation},
\end{align} 
from which we identify the critical value of $h$ as \mbox{$h_c=1/2$}. This field strength marks the transition between the paramagnetic ($|h|>h_c$) and ferromagnetic ($|h|<h_c$) phases. For the energy per spin we find
\begin{align}
	\lim_{N \rightarrow \infty} \frac{\expval{H}}{N}  &= \begin{cases}
		-\frac{1}{2}(h^2+\frac{1}{4}) & |h|\leq h_c\\
		-\frac{|h|}{2} & |h|>h_c
	\end{cases}.\label{eq:LMG_Energy}
\end{align}  
Both Eqs.~\eqref{eq:LMG_Magnetisation} and \eqref{eq:LMG_Energy} are exact results for the thermodynamic limit.

For finite systems, the optimal value of $\theta$ is non-zero, and the layer of $C^\theta R^\theta$ rotations in Eq.~\eqref{eq:ansatz} will introduce correlations between the spins. This brings about a major improvement in accuracy compared to the product state mean-field ansatz. We further this improvement by restoring in our ansatz the symmetries present in the LMG Hamiltonian \eqref{eq:lmg_hamiltonian}. Specifically, $H$ exhibits permutation symmetry under the exchange of any two spins, and also parity symmetry under a $\pi$-rotation about the $x$-axis, which sends $(X_i,Y_i,Z_i)$ to $(X_i,-Y_i,-Z_i)$. We enforce these symmetries on the ansatz $\ket{\psi_t}$ by projecting it onto the relevant symmetry subspaces. As shown in Section III.C of the SM \cite{lourensSupplementaryMaterial}, this yields a state $\ket{\psi_s}$ within the $(2S+1)$-dimensional subspace corresponding to the maximum magnitude $S=N/2$ of the total spin. The analytic expression for $\ket{\psi_s}$, parametrised by $\theta$ and $\phi$, now serves as a refined version of the original ansatz. We use this symmetrised ansatz to estimate the ground-state energy as well as the RMS magnetisation
\begin{equation}
	M_{\rm rms}=\frac{1}{2N}\sqrt{\left\langle\left({\textstyle \sum}_i Z_i\right)^2\right\rangle}.
	\label{eq:RMS_Magnetisation}
\end{equation}
\begin{figure}[t!]
    \centering
    \includegraphics[width=\linewidth]{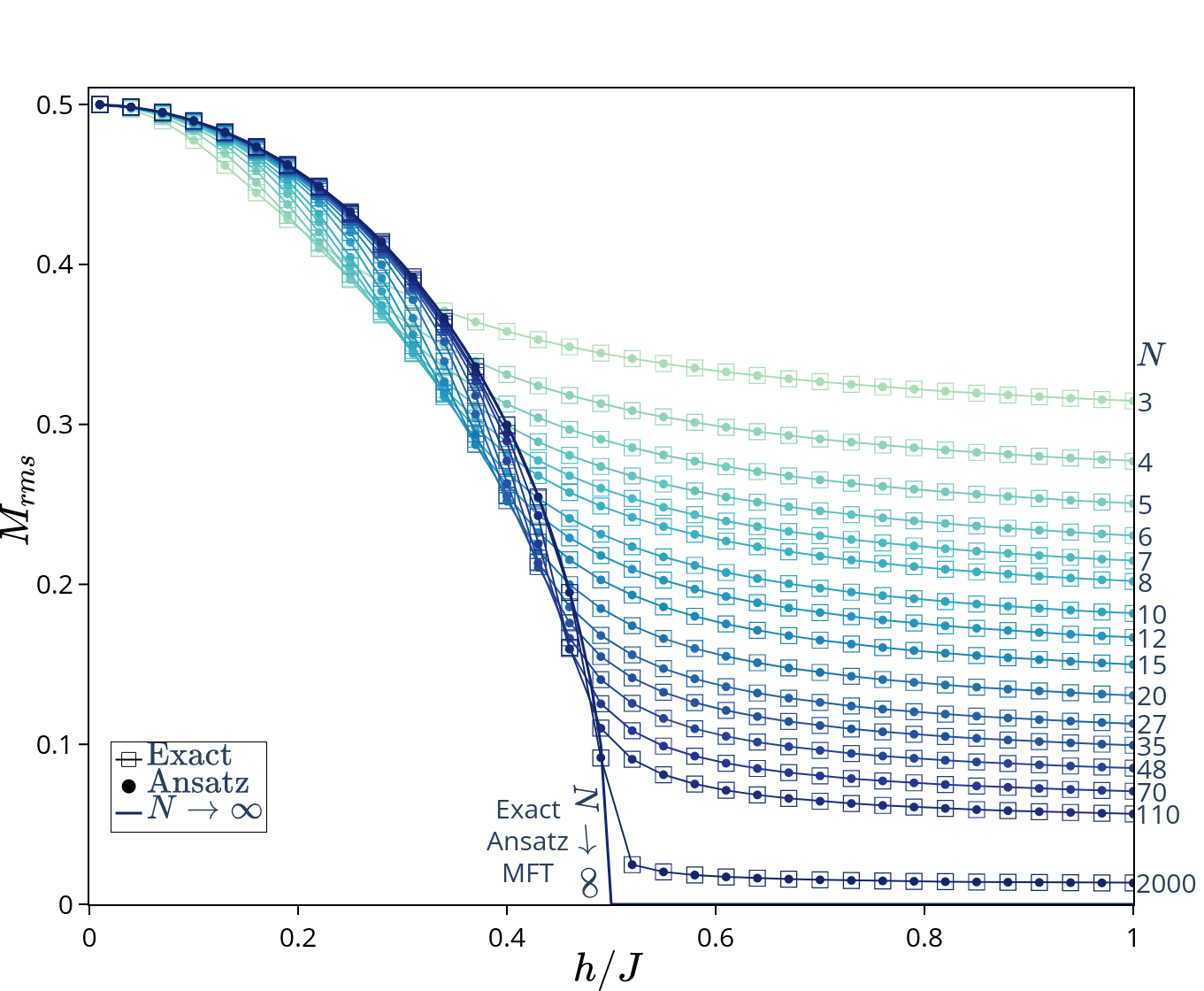}
    \vspace{-13pt}
    \caption{RMS magnetisation of Eq.~\eqref{eq:RMS_Magnetisation} vs $h/J$ for the LMG model. Exact results compared to the symmetrised ansatz.}
    \vspace{-12pt}
    \label{fig:lmg_results0}
\end{figure}
\begin{figure}[t!]
    \centering
    \includegraphics[width=\linewidth]{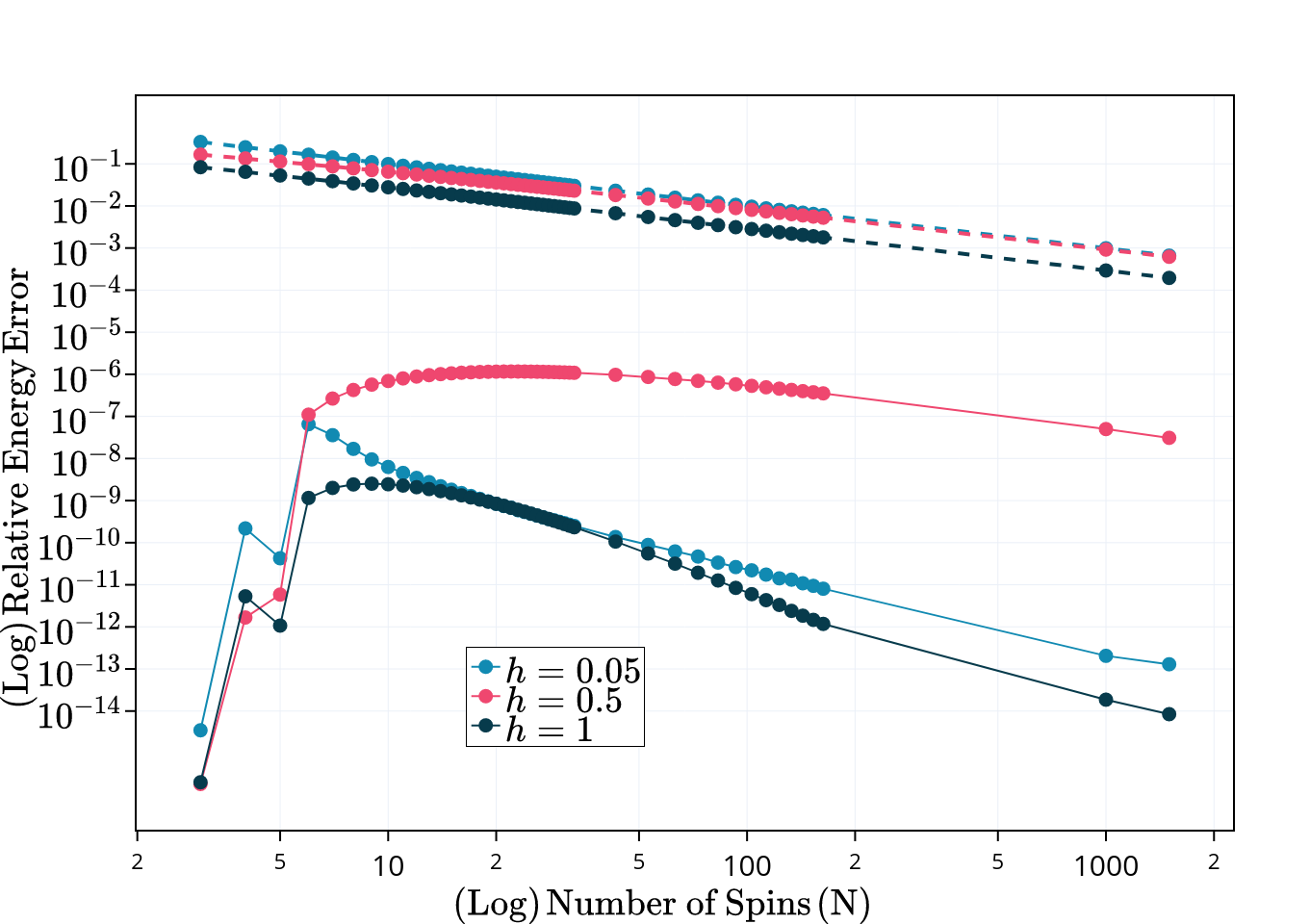}
    \vspace{-13pt}
    \caption{Relative ground-state energy error vs $N$ for the LMG model at different field strengths $h$. Compares symmetrised ansatz (solid) and MFT (dashed).}
    \vspace{-14pt}
    \label{fig:lmg_results1}
\end{figure}
Fig.~\ref{fig:lmg_results0} shows the result of this calculation of $M_{\rm rms}$ for different numbers of spins $N$. Remarkably, there is no visible difference between our ansatz-based result and the exact value of the magnetisation. This suggests that the symmetrised ansatz $\ket{\psi_s}$ captures finite-size effects very accurately. Fig.~\ref{fig:lmg_results1} shows the relative error in the ground-state energy, $\epsilon_{\rm{rel}} = \lvert E_{\rm{pred}} - E_{\rm{exact}}\rvert /\lvert E_{\rm{exact}}\rvert$, for different field strengths $h$, plotted on a logarithmic scale as $N$ increases. For our symmetrised ansatz, this error is at most of order $10^{-6}$ for $h=h_c=1/2$ and about $N=25$, and tends to zero as $N$ increases. The result of using the mean-field product state ansatz (with $\theta=0$) is also shown. While this too becomes exact in the thermodynamic limit, it fares much worse than the symmetrised ansatz for finite system sizes. Both the translationally invariant ansatz $\ket{\psi_t}$ and its symmetrised counterpart $\ket{\psi_s}$ produce exact results for $M_{\rm rms}$ and the ground-state energy within the thermodynamic limit. However, these quantities probe limited features of the two states, and it turns out that $\ket{\psi_t}$ and $\ket{\psi_s}$ have fundamentally different characters, even in this limit. Specifically, we found that $\ket{\psi_t}$ reduces to a mean-field product state as $N\rightarrow\infty$ due to the optimal value of $\theta$ vanishing. In contrast, optimising $\ket{\psi_s}$ yields a non-zero $\theta$, even in the $N\rightarrow\infty$ limit, thereby retaining the entanglement from the $C^\theta R^\theta$ rotations.\clearpage
\onecolumngrid
\begin{figure*}[!t]
    \centering
    \includegraphics[width=\textwidth]%
                    {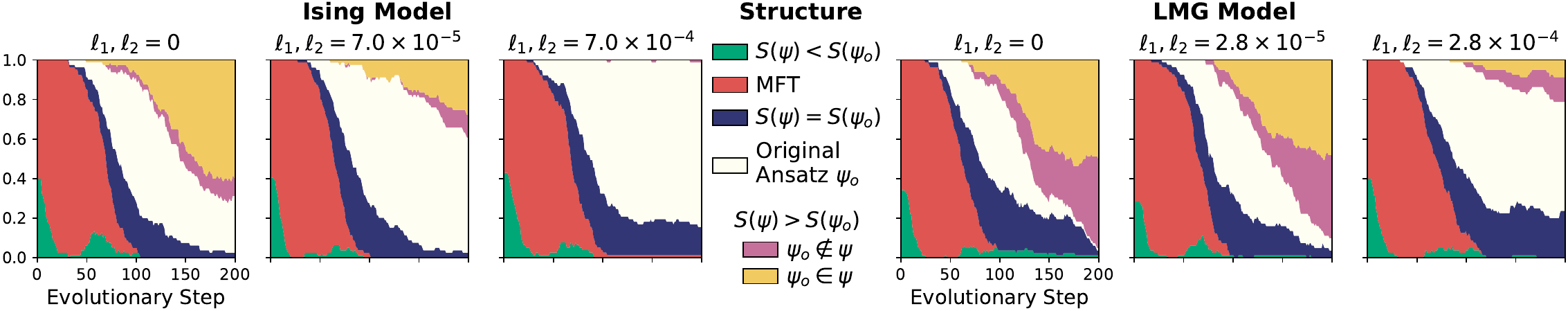}
            \caption{The density of particular structural classes of discovered ansatzes at different evolutionary steps. Each subplot shows results for a different complexity penalty pair $(l_1,l_2)$(increasing to the right). At each step (vertical slice) the lengths of intervals of a particular colour indicate the fraction of runs which yielded a fittest individual within a specific class.  The classes are determined by the structural complexity $S(\psi)$ of the fittest individual $\psi$. White represents the original ansatz $\psi_o$ in Eq.~\eqref{eq:ansatz}. Green and Red denote ansatzes with structural complexity less than the original, with Red being the mean‑field solution. Blue indicate ansatzes with structural complexity equal to the original. Pink and Yellow are more complex: yellow contains the original ansatz as a building block, while pink does not. }
            \vspace{-20pt}
    \label{fig:structure_densities}
\end{figure*}
\twocolumngrid
\noindent This aligns with Refs.~\cite{dusuelFiniteSizeScalingExponents2004,barthelEntanglementEntropyFree2006,orusEquivalenceCriticalScaling2008} where it is shown that the \emph{exact} ground state in the paramagnetic phase always contains non-trivial entanglement and does not reduce to a product state as $N\rightarrow\infty$. These observations underscore that the symmetrisation step can fundamentally alter the correlations present in the ansatz, and that optimising the variational parameters before versus after this step can yield very different results.\\\vspace{-6pt}

\PSect{Robustness of the search} Fig.~\ref{fig:structure_densities} shows how varying the complexity penalties $l_1$ and $l_2$ in the fitness function \ref{eq:fitness} affects the search outcome. Each subplot contains data from $79\pm(3)$ runs where the selection pressure and exploration parameters are fixed at $\rho=0.01$ and $\epsilon=0.33$, and where the fitness is evaluated for the set of system sizes $\mathcal{N}=\{5,6,7\}$. The fittest individual at each evolutionary step is classified into one of six classes indicated by colour. The classes are based on the structural complexity of that individual $S(\psi)$ relative to that of the original ansatz $S(\psi_o)$: green and red denote lower complexity, blue and white equal complexity, and pink and yellow higher complexity. Between the mean-field (red) and original ansatz (white) there are competing local minima (blue) with exactly the same structural complexity as the original, namely $S(\psi_o^{(n)})=3$ for all $n\in\mathcal{N}$ but with lower fitness. Most prominent in this class was the ansatz
\begin{align}
    \ket{\psi_x}&= \left(\overleftarrow{\prod_{k=0}^{N-1}}  R^\phi_k \right)  \left( \overleftarrow{\prod_{j=0}^{N-1}}  e^{-i\frac{\theta}{2}X_{j}Y_{j+1}}\right) \ket{z,+}^{\otimes N}, \label{eq:ansatz_xy}
\end{align}
with $j+N\equiv j$. We see that for the largest penalties $l_1$ and $l_2$, the fittest individual after $200$ steps was the original ansatz in $83\%$ and $55\%$ of runs for the Ising model and LMG model respectively. As we decrease the complexity penalties more complex structures begin to emerge, shown in yellow and pink. These structures have $S(\psi)>3$, and are distinguished by whether they contain the original ansatz as a building block (yellow) or not (pink). The prevalence of the former yellow class for lower complexity penalties shows that the original ansatz is a robust structure that functions as a building block for more complex ansatzes. Fig.~\ref{fig:structure_densities} also shows that the effect of the complexity penalties is model dependent. With decreasing complexity penalties the LMG model is dominated by complex structures more readily than the Ising model. This results from the smaller energy difference found between different network structures for the LMG model. Still, the original ansatz (white) is a prominent beyond-mean-field structure in all cases.\\\vspace{-6pt}

\PSect{Conclusion}We have introduced a general method for constructing ground-state ansatzes that are both analytically tractable and quantitatively accurate across a wide range of system parameters. Our approach can be applied to any physical system that is amenable to a variational treatment in terms of tensor network states. For example, finding time evolution approximations would entail a train-by-example strategy. The subsequent work~\cite{rouillardAutomatedQuantumAlgorithm2025} use our framework with this strategy and shows how to automatically generate the $N$-qubit realisations of popular quantum algorithms (Deutsch–Jozsa, QFT, Grover) by considering only examples up to five qubits. The core of our approach lies in the interplay between the domain-specific language and the fitness criteria. The former enables fitness evaluation on small system sizes, which is computationally efficient and allows capturing system-size scaling. The latter favours ansatzes with low variational and structural complexity while preserving accuracy. This results in expressive ansatzes which tend to break the underlying model's symmetries, but due to their simple structure, these symmetries can be restored analytically. This provides a systematic way to improve the ansatz.\\
Remarkably, by applying our method to both the LMG and TFIM models, the algorithm autonomously constructs a mean-field treatment and extends it to incorporate correlations. This provided us with a simple and interpretable structure. For the LMG model it yields highly accurate results for finite systems, far surpassing that of a mean-field treatment, and which become exact in the thermodynamic limit. For the TFIM case discussed in the end matter, we obtain accurate results across all system sizes and greatly improve upon the mean-field treatment in the thermodynamic limit. 

\PSect{Acknowledgements}The authors thank Amy Rouillard for valuable discussions. Support from the Oppenheimer Memorial Trust, the Department of Science, Technology and Innovation, and the National Research Foundation of South Africa is kindly acknowledged. Support for the Python package development by the Unitary Foundation is also kindly acknowledged. The authors acknowledge the Centre for High Performance Computing (CHPC),
South Africa, for providing computational resources to this research project.

\bibliography{refs_override,references}
\clearpage
\onecolumngrid
%---------------------------------------------------------------------------
\begin{center}
\ \vskip 0.0cm
{\large\bf End Matter}
\end{center}
% \begin{figure}[h]
%     \centering
%     \placeholderrect[gray]{.9\textwidth}{0.2\textheight}
%     \caption{(Placeholder) Phase diagram overview.}
%     \label{fig:A1-placeholder}
% \end{figure}
% \begin{figure}[h]
%     \centering
%     \includegraphics[width=\textwidth]%
%                     {figures/structure_densities_side_by_side.pdf}
%             \caption{Density of discovered ansatz structures per evolutionary step for the Ising (left) and LMG (right) models. Colours denote structure classes, ordered first by complexity and then fitness—thus the most complex/fittest appear near the top. Red is the mean‑field (MFT) solution; white is the original ansatz. Green, dark blue and light‑pink share the same structural complexity as the original (white), which attains the best fitness among them. Dark pink and yellow are more complex: yellow contains the original ansatz as a building block, pink does not. Each subplot shows a different regularisation pair $(\ell_1,\ell_2)$ for structural/variational complexity; moving right increases the penalty.}
%     \label{fig:structure_densities}
% \end{figure}
%---------------------------------------------------------------------------
\twocolumngrid
\makeatletter
\setlength{\@dblfptop}{0pt}                 % no extra space above dbl top floats
\renewcommand{\dbltopfraction}{0.99}        % allow big dbl floats at top
\renewcommand{\textfraction}{0.01}          % let pages be mostly float
\makeatother
%---------------------------------------------------------------------------

\begin{figure}[htb]
    \centering
    \includegraphics[width=\linewidth]%
                    {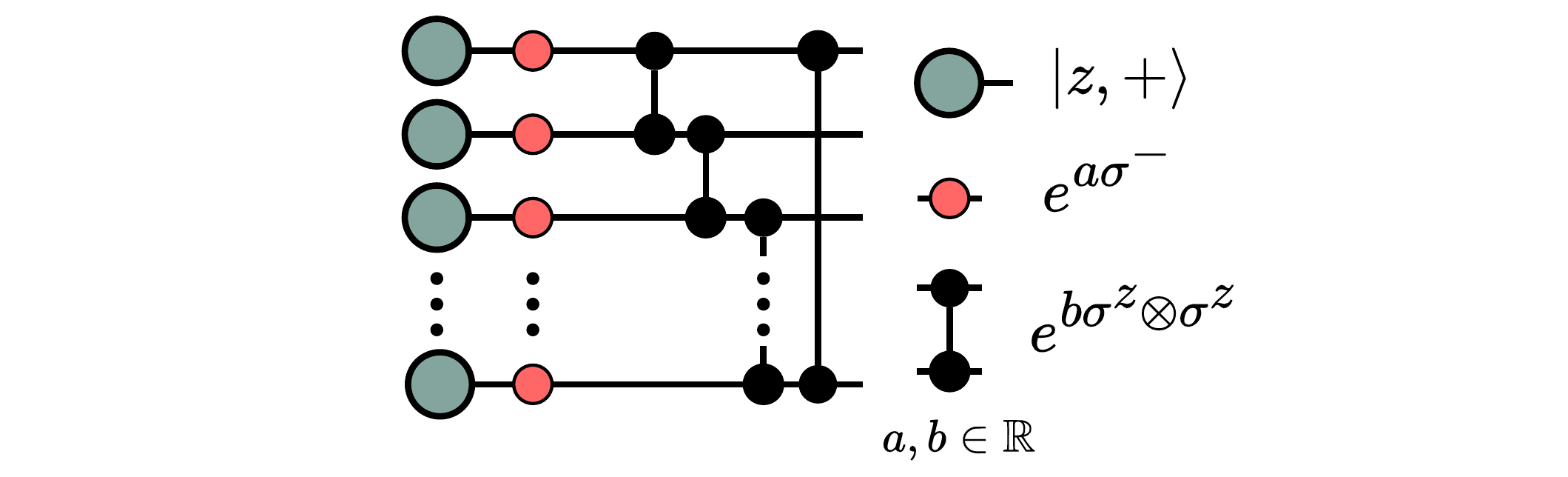}
    \caption{Ansatz discovered with the ladder‑operator basis $\{I, \sigma^+, \sigma^-, Z\}$. The search returns a translationally invariant, two‑parameter periodic MPS Eq.~\eqref{eq:mzz_fix_trans_mps}, and it attains the same energy expectation values as the original ansatz in Fig.~\ref{fig:ansatz}.}
    \label{fig:ladder_anz}
\end{figure}

%%%%%%%%%%%%%%%%%%%%%%%%%%%%%%%%%%%%%%%%%%%%%%%%%%
\renewcommand{\theequation}{A\arabic{equation}}
%%%%%%%%%%%%%%%%%%%%%%%%%%%%%%%%%%%%%%%%%%%%%%%%%%
\setcounter{equation}{0}
\PSect{Results for the TFIM}The Hamiltonian for the TFIM with $N$ spin-$\frac{1}{2}$ particles on a periodic chain is
\begin{align}
    H&=-\frac{J}{4}\sum_{i=0}^{N-1} Z_iZ_{i+1}-\frac{h}{2}\sum_{i=0}^{N-1} X_i,\label{eq:tfim_hamiltonian}
\end{align}
with $J$ and $h$ again the interaction and external field strengths. We set $J=1$ as before.
For the LMG model it was seen that the mean-field ansatz with $\theta=0$ in Eq.~\eqref{eq:ansatz} was sufficient for correctly predicting the critical value of the external field strength and for calculating the order parameter $M_{\rm rms}$ in the thermodynamic limit.
%For the LMG model it was seen that the mean-field treatment becomes exact in the thermodynamic limit, and that the second layer of $C^\theta R^\theta$ rotations in Eq.~\eqref{eq:ansatz} therefore only served to improve the accuracy of the ansatz for finite systems. 
For the TFIM the situation is quite different. Here, even in the thermodynamic limit, the $C^\theta R^\theta$ rotations play a crucial role in introducing correlations between spins, and are essential for shifting the estimate for the critical field strength closer to its true value. Using the ansatz in Eq.~\eqref{eq:trans_mps} together with Eqs.~\eqref{eq:ti_field_mag} - \eqref{eq:ti_interaction_corr} we find the energy per spin in the thermodynamic limit to be
\begin{align}
    \lim_{N\rightarrow \infty}\frac{\expval{H}}{N} &= -\frac{s-t}{4(st-1)}\left ( (s-t)+2hc^2\right )-\frac{1}{4}.
\end{align}
\begin{figure}[t]
    \centering
    \includegraphics[width=\linewidth]{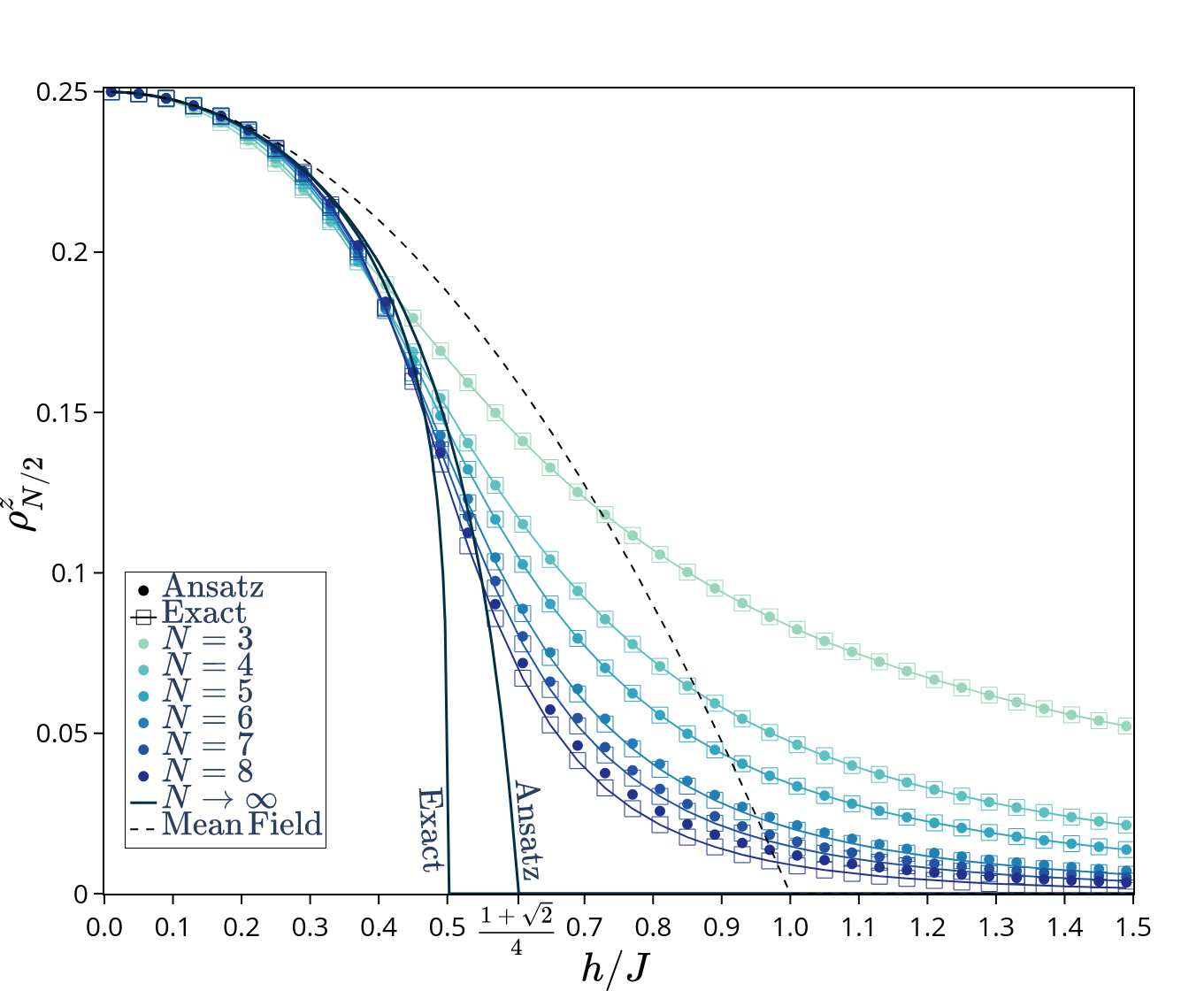}
    \caption{The long-range correlation function $\rho^z_{N/2}$ versus $h/J$ for the TFIM. Results are shown for the symmetrised ansatz (finite $N$, $N\rightarrow\infty$), together with exact values and the mean-field prediction.}
    \label{fig:ising_results0}
\end{figure}
By minimising this expression with respect to $\theta$ and $\phi$ we identify a critical field strength of 
\begin{align}
	h_c&=\frac{1+\sqrt{2}}{4}\approx 0.604,
\end{align}
above which the magnetisation $\expval{Z_i}$ in Eq.~\eqref{eq:ti_int_mag} vanishes. See Section IV of the Supplementary Material the details of this calculation.  This estimate for $h_c$ is indeed closer to the exact value $h^{\rm ex}_c=0.5$ when compared to the mean-field result of $h^{\rm mf}_c=1$, which would follow from setting $\theta=0$ and only varying $\phi$.
While the TFIM Hamiltonian lacks the permutation symmetry of the LMG model, it retains the parity symmetry. We again restore this symmetry by projecting the ansatz Eq.~\eqref{eq:trans_mps} onto the positive symmetry subspace to produce a modified ansatz $\ket{\psi_p}$. See Section II.D of the SM \cite{lourensSupplementaryMaterial}. Using $\ket{\psi_p}$ we calculate the long-range correlation function $\rho^z_{N/2}=\frac{1}{4}\expval{\psi_p|Z_0Z_{N/2}|\psi_p}$, which serves as an order parameter for characterising the model's two phases. Fig.~ \ref{fig:ising_results0} shows the result this calculation for various system sizes. While our ansatz-based result matches the exact one closely for small $N$, it begins to deviate from it as $N$ increases. This is to be expected due to the error in the ansatz's prediction of the critical field strength. The mean-field result, with its prediction of $h^{\rm mf}_c=1$, is also shown.\\
%  \hk{Is die volgende nodig hier? Kan ons nie na die SM verwys nie?} When $|h|>h_c$ the optimal values of $s=\sin(\theta)$ and $t=\sin(\theta+\phi)$ are found to be $s=(4h)^{-1}$ and $t={\rm sgn}(h)$, while for $|h|<h_c$ these need to be solved from 
% \begin{align}
% 	h=\frac{2s}{(s^2+1)^2}\qquad\text{and}\qquad t=2hs^2+2h-s.
% \end{align}
\PSect{Other choices of operator basis}
The original ansatz from Eq.~\eqref{eq:ansatz} is represented as a product of unitary operators, meaning it is equivalent to a quantum circuit. This is because the initial operator pool was generated from the Pauli-matrices $\sigma\in \{I,X,Y,Z\}$ in the form $e^{i\phi \sigma_i},e^{i\theta \sigma_i\otimes\sigma_j}$ where $\theta,\phi \in \mathbb{R}$. We can change the operator pool to include non-unitary tensors to find more general tensor networks. For example, using an operator basis that contains ladder operators, i.e. $\{I, \sigma^+, \sigma^-, Z\}$ our method finds another efficient ansatz shown in Fig.~\ref{fig:ladder_anz}. 
With this operator basis the search was able to obtain a translationally invariant ansatz. Numerically, this ansatz yields the same energy expectation values as the original ansatz in Fig.~\ref{fig:ansatz} for all system sizes. We can express this ansatz as the following two-parameter periodic MPS
\begin{equation}
    \ket{\psi_l} = \frac{1}{M_l} \sum_{\vec{\sigma}} {\rm Tr}\bigl(C^{\sigma_0} C^{\sigma_1} \cdots C^{\sigma_{N-1}}\bigr) \ket{z,\vec{\sigma}}
    \label{eq:mzz_fix_trans_mps}
\end{equation}
where $M_l$ is a normalisation factor and the matrices $C^{\pm}$ are given by
\begin{align}
    C^{\pm} &=a^{\frac{1\mp 1}{2}}
    \begin{bmatrix}
        \cosh(b) & \pm\cosh(b) \\
        \pm\sinh(b) & \sinh(b)
    \end{bmatrix},
\end{align}
with $a,b\in\mathbb{R}$. Expectation values may be obtained in the same fashion as detailed for the original ansatz in Section II of the SM \cite{lourensSupplementaryMaterial}. The choice of operator basis will depend on the context of the problem. For example, if the goal is to obtain short-depth quantum circuits for state preparation on a specific quantum device, then using native gates for that hardware will be beneficial.\\

\begin{figure}[t]
    \centering
    \includegraphics[width=\linewidth]{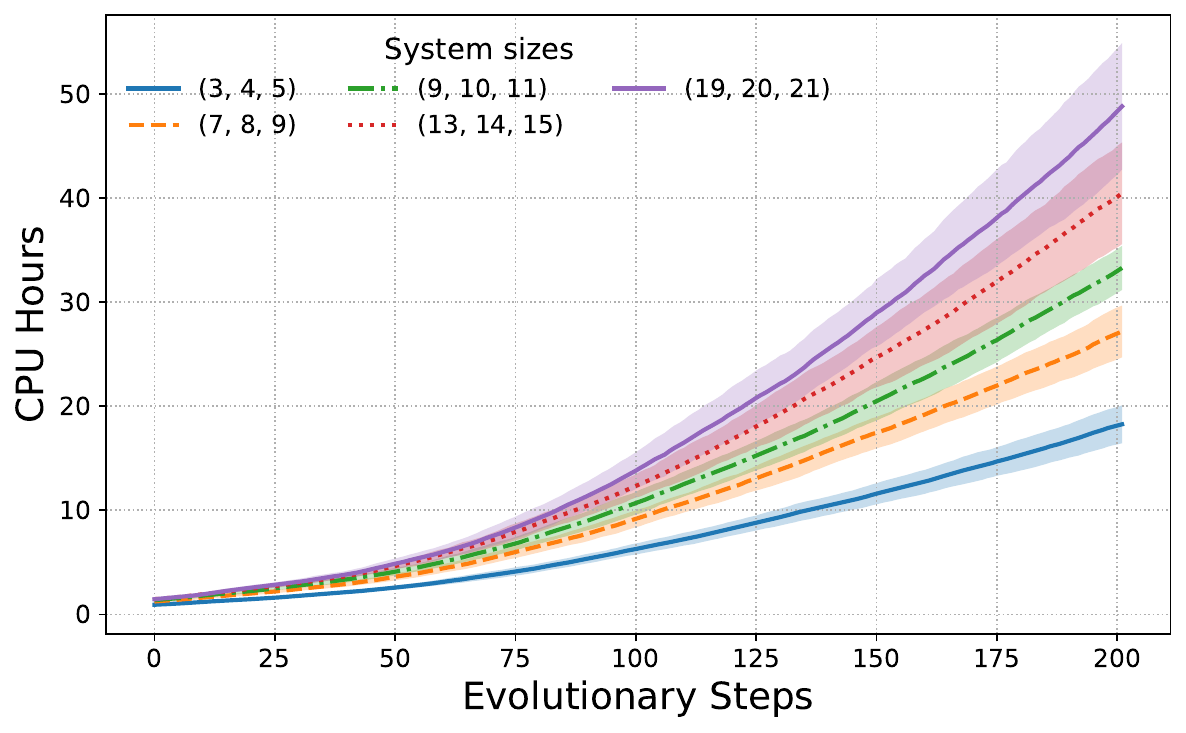}
    \caption{Runtime of the evolutionary search for the Ising model. We consider five system–size groups, each averaged over 15 runs (75 runs total), using complexity penalties $l_1=l_2=7\times10^{-4}$ and selection/exploration parameters $\rho=0.01$ and $\epsilon=0.33$.}
    \label{fig:evo_runtime}
\end{figure}

\begin{figure}[t]
    \centering
    \includegraphics[width=\linewidth]{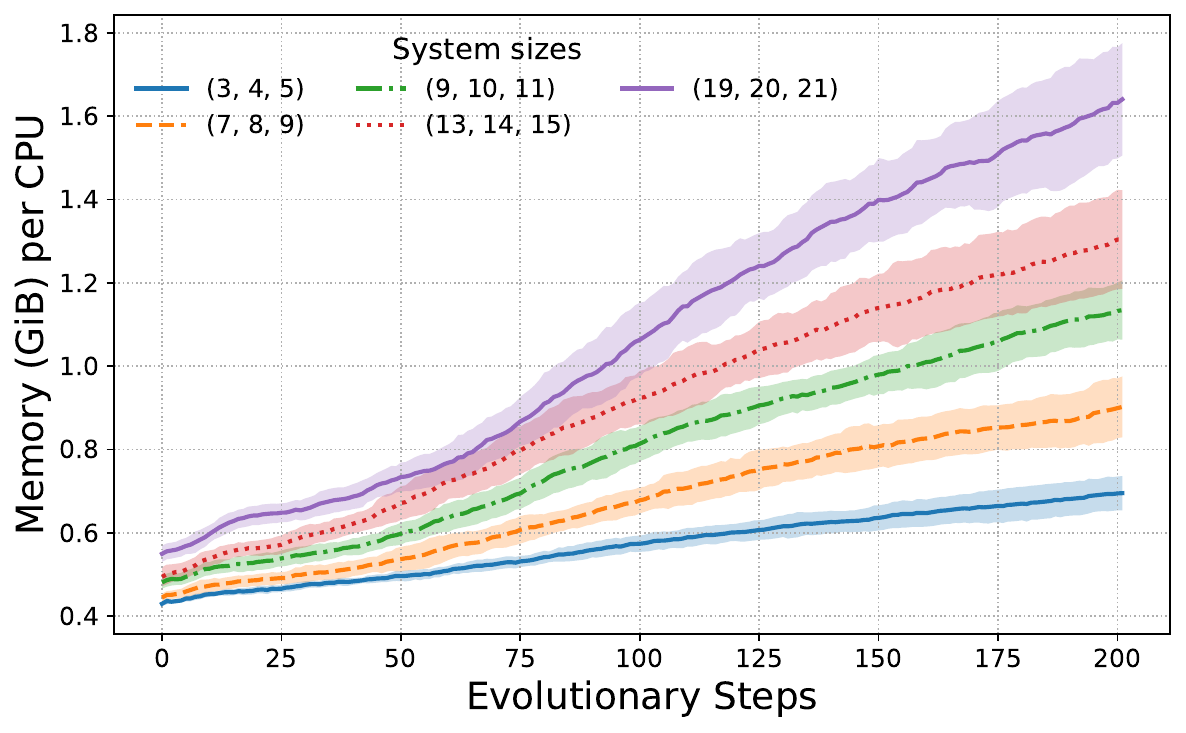}
    \caption{Per‑CPU memory usage during evolutionary search. Fitness evaluations are run in parallel, so each CPU corresponds to one worker. Memory therefore depends on both the number of workers and the complexity penalty.}
    \label{fig:evo_memusage}
\end{figure}
\PSect{Computational cost} The main bottleneck in the evaluation of individuals during the evolutionary search is memory, as the bond dimension between two sites grows exponentially with the number of tensors that connect them. Increasing the complexity penalty limits the exploration of overly large network structures and, in turn, allows larger system sizes to be considered. Because the DSL automatically scales over system sizes, it can explore structures that, in principle, extend well beyond what we can evaluate explicitly for large systems. Figs.~\ref{fig:evo_runtime} and \ref{fig:evo_memusage} report the computational cost of our method over 75 searches for an Ising‑model ansatz, organised into five size groups and averaged over 15 runs each. We used $l_1=l_2=7\times10^{-4}$, $\rho=0.01$, and $\epsilon=0.33$ for all runs. In 65 of the 75 cases ($86\%$), the fittest individual at step 200 was the original ansatz, close to the 83\% ratio observed at step 200 for system sizes $n\in\{5,6,7\}$ in Fig.~\ref{fig:structure_densities} with the same penalty. This suggests that the converged ansatz is largely insensitive to the system sizes considered. We evaluated system sizes up to $n=21$ spins, enabled by the complexity penalty. From Fig.~\ref{fig:structure_densities}, beyond-mean‑field structures begin to dominate around 100 steps; in Fig.~\ref{fig:evo_runtime} this corresponds to $6.27$ CPU‑hours on average for the smallest group, $n\in\{3,4,5\}$, and about $13.8$ CPU‑hours for the largest, $n\in\{19,20,21\}$. Fig.~\ref{fig:evo_memusage} shows the per‑worker memory footprint. For $100$ evolutionary steps and $n\in\{3,4,5\}$ we required less than 0.57\,GiB per worker on average, with each worker occupying one CPU core. For $100$ steps and $n\in\{19,20,21\}$ each worker required less than 1.06\,GiB on average.\\\\
After finding an ansatz with the search algorithm, we optimise its variational parameters for each system size. This optimisation may not always be required, for example with the Ising model the optimal angles essentially become constant once $N>20$. Therefore, re-optimisation for larger system sizes is not required. For the LMG model, the situation is different. For the ferromagnetic phase, numerical results suggest that the leading-order finite-size corrections scale like $1/N$. Using only system sizes $N\in\{50,100,150\}$, we can extrapolate to arbitrary system sizes without losing accuracy. For the paramagnetic phase, extrapolation to the thermodynamic limit is no longer reliable. The symmetrised ansatz produces essentially exact results in this phase, which might cause the optimal parameter values to depend more sensitively on the system size.\\\\
\PSect{Outlook} In addition to the two models studied in the main text, we are in the process of applying our approach to the quantum axial next-nearest-neighbour Ising (ANNNI) model. This non-integrable model exhibits a rich phase structure, and it might prove to be beneficial to perform independent searches within the different phases. Interestingly, preliminary results indicate that we are able to obtain simple, interpretable structures that perform well within a particular phase, but break down sharply near the phase boundaries. This suggests that the physics of the different phases could be probed by investigating changes in the structure of the ansatz from one phase to another.

%---------------------------------------------------------------------------
\clearpage
\onecolumngrid
\renewcommand{\theequation}{S\arabic{equation}}
\renewcommand{\thefigure}{S\arabic{figure}}
%---------------------------------------------------------------------------
\setcounter{equation}{0}

\section{MPS Derivation}
Here we show how the MPS form of our ansatz in Eq.~$(4)$ of the main text can be derived from its original form in Eq.~$(2)$. Recall that the ansatz generated by our evolutionary algorithm produces, for $N$ spins, the state
\begin{align}
    \ket{\psi_o} &= \left(\overleftarrow{\prod_{k=0}^{N-1}}  C^{\theta}_{k,k+1} R^{\theta}_{k+1} \right)  \left( \overleftarrow{\prod_{j=0}^{N-1}} R^\phi_j \right) \ket{z,+}^{\otimes N}, \label{eq:supp_anzcircuit}
\end{align}
where $\phi$ and $\theta$ are variational parameters and $\ket{z,\pm}$ are the eigenstates of the Pauli-$Z$ matrix with eigenvalues $\pm 1$. The two unitary operators appearing in $\ket{\psi_o}$ are
\begin{align}
	C^{\theta}_{ij} = e^{i\frac{\theta}{2} Z_i Y_j}\qquad\text{and}\qquad R^\theta_j = e^{-i\frac{\theta}{2} Y_j}.\label{eq:supp_unitaryops}
\end{align}
All site indices are treated as periodic, i.e. $i\equiv i+N$. The operator products in Eq.~\eqref{eq:supp_anzcircuit} are ordered with the $k$ and $j$ indices decreasing from left to right. For example, if $N=3$ the state reads
\begin{align}
    \ket{\psi_o} &= C^{\theta}_{2,0} R^{\theta}_{0}  C^{\theta}_{1,2} R^{\theta}_{2} C^{\theta}_{0,1} R^{\theta}_{1} R^\phi_2 R^\phi_1 R^\phi_0  \ket{z,+}_0\otimes\ket{z,+}_1\otimes\ket{z,+}_2.
\end{align}
It will be convenient to perform this derivation in the $Y$ basis due to the multiple Pauli-$Y$ operators appearing in Eq.~\eqref{eq:supp_anzcircuit}. We therefore introduce
\begin{align}
    \ket{\sigma_k} &\equiv \ket{y,\sigma_k},
\end{align}
with $\sigma_k=\pm$ to denote the eigenstates, with eigenvalues $\pm1$, of the Pauli-Y matrix associated with the spin at site $k$. The initial product state from Eq.~\eqref{eq:supp_anzcircuit} can be expressed in this basis as
\begin{align}
    \ket{z,+}^{\otimes N} &= \nf \sum_{\vec{\sigma}} \ket{\vec{\sigma}}.
\end{align}
We will recast Eq.~\eqref{eq:supp_anzcircuit} in MPS form by successively applying the operator blocks $C^{\theta}_{k,k+1} R^{\theta+\phi}_{k+1}$ to the initial product state $\ket{z,+}^{\otimes{N}}$. Let 
\begin{align}
    U(k_0)&= \left(\overleftarrow{\prod_{k=k_0}^{N-1}}   C^{\theta}_{k,k+1} R^{\theta+\phi}_{k+1} \right)
\end{align}
denote the final $N-k_0$ blocks to be applied. Since $C^\theta_{ij}$ commutes with $R^\phi_k$ when $i\neq k$, the form of $\ket{\psi_o}$ in Eq.~\eqref{eq:supp_anzcircuit} can be written as 
\begin{align}
    \ket{\psi_o} &= R_0^{-\phi}U(0)R_0^{\phi}\ket{z,+}^{\otimes N}=\nf  R_0^{-\phi}U(0)R_0^{\phi}\sum_{\vec{\sigma}} \ket{\vec{\sigma}}.\label{eq:supp_anzdoubleangle}
\end{align}
The action of the two operators in Eq.~\eqref{eq:supp_unitaryops} on the $\ket{\sigma}$ states are
\begin{equation}
    C^{\theta}_{ij}\ket{\sigma_i\sigma_j}=c\ket{\sigma_i\sigma_j}+is\sigma_j\ket{-\sigma_i\sigma_j}\qquad{\rm and}\qquad
    R^{\theta}_{j}\ket{\sigma_j}=e^{-i\frac{\theta}{2}\sigma_j}\ket{\sigma_j},
\end{equation}
where 
\begin{equation}
	c\equiv\cos\Bigl({\frac{\theta}{2}}\Bigr)\qquad{\rm and}\qquad s\equiv\sin\Bigl({\frac{\theta}{2}}\Bigr).
\end{equation}
Using the actions above, we begin to rewrite $\ket{\psi_o}$ by applying $R_0^\phi$ together with the first (rightmost) block of operators $C_{01}^{\theta}R^{\theta+\phi}_1$ in $U(0)$ to the initial state:
\begin{align}
    \ket{\psi_o} &=\nf R_0^{-\phi}U(0)R_0^{\phi} \sum_{\vec{\sigma}} \ket{\vec{\sigma}}\\
    &=\nf R_0^{-\phi}U(1) \sum_{\vec{\sigma}} C_{01}^{\theta}R^{\theta+\phi}_1R^\phi_0\ket{\vec{\sigma}}\\
    &=\nf R_0^{-\phi}U(1) \sum_{\vec{\sigma}} e^{-i\frac{(\theta+\phi)}{2}\sigma_1}e^{-i\frac{\phi}{2}\sigma_0} \left (c\ket{\sigma_0\sigma_1}+is\sigma_1\ket{-\sigma_0\sigma_1}\right )\otimes\ket{\vec{\sigma}^\prime}\\
    &=\nf R_0^{-\phi}U(1) \sum_{\vec{\sigma}} e^{-i\frac{(\theta+\phi)}{2}\sigma_1} \left (ce^{-i\frac{\phi}{2}\sigma_0}+is\sigma_1e^{i\frac{\phi}{2}\sigma_0}\right )\ket{\vec{\sigma}}.
\end{align}
Notice how we can write the amplitude inside the sum as an inner product of vectors as 
\begin{equation}
    e^{-i\frac{(\theta+\phi)}{2}\sigma_1} \left (ce^{-i\frac{\phi}{2}\sigma_0}+is\sigma_1e^{i\frac{\phi}{2}\sigma_0}\right )=\begin{bmatrix}e^{-i\frac{\phi}{2}\sigma_0}&e^{i\frac{\phi}{2}\sigma_0}\end{bmatrix}  \begin{bmatrix}ce^{-i\frac{(\theta+\phi)}{2}\sigma_1}\\is\sigma_1e^{-i\frac{(\theta+\phi)}{2}\sigma_1}\end{bmatrix} \equiv  \mathbf{b}(\sigma_0)^{T}\mathbf{a}({\sigma_1}), \label{eq:supp_ba_def}
\end{equation}
which also serves to define the vectors $\mathbf{a}(\sigma_i)$ and $\mathbf{b}(\sigma_i)$.
After the application of the first operator block in $U(0)$, $\ket{\psi_o}$ therefore reads
\begin{align}
    \ket{\psi_o} &= R_0^{-\phi}U(1)\nf\sum_{\vec{\sigma}} \mathbf{b}(\sigma_0)^{T}\mathbf{a}({\sigma_1})\ket{\vec{\sigma}}.
\end{align}
It follows via induction that after applying the rightmost $k$ operator blocks from $U(0)$, the resulting form of $\ket{\psi_o}$ will be
\begin{align}
	\ket{\psi_o} &=R_0^{-\phi} U(k)\nf\sum_{\vec{\sigma}} \mathbf{r}(\sigma_0,\dots,\sigma_{k-1})^{T} \mathbf{a}({\sigma_k})  \ket{\vec{\sigma}},\label{eq:psi0intermsofr}
\end{align}
where $\mathbf{r}(\sigma_0,\dots\sigma_{k-1})^{T}$ is a row vector depending on the spin states $\sigma_0,\dots,\sigma_{k-1}$. To prove this claim and obtain an expression for $\mathbf{r}(\sigma_0,\dots,\sigma_{k-1})$, we start from Eq.~\eqref{eq:psi0intermsofr} and write
\begin{align}
    \ket{\psi_o} &=\nf  R_0^{-\phi}U(k) \sum_{\vec{\sigma}} \mathbf{r}(\sigma_0,\dots,\sigma_{k-1})^{T} \mathbf{a}({\sigma_k})  \ket{\vec{\sigma}}\\
    &=\nf R_0^{-\phi}U(k+1) \sum_{\vec{\sigma}}  \mathbf{r}(\sigma_0,\dots,\sigma_{k-1})^{T}\mathbf{a}({\sigma_k}) \left ( C_{k,k+1}^{\theta}R^{\theta+\phi}_{k+1} \ket{\vec{\sigma}} \right ) \\
    &=\nf R_0^{-\phi}U(k+1) \sum_{\vec{\sigma}}  \mathbf{r}(\sigma_0,\dots,\sigma_{k-1})^{T}\mathbf{a}({\sigma_k})  \left (e^{-i\frac{(\theta+\phi)}{2}\sigma_{k+1}}\right ) \Bigl ( c \ket{\sigma_k\sigma_{k+1}}+is\sigma_{k+1}\ket{-\sigma_k\sigma_{k+1}}\Bigr )\otimes\ket{\vec{\sigma}^\prime} \\
    &=\nf  R_0^{-\phi}U(k+1) \sum_{\vec{\sigma}}   \mathbf{r}(\sigma_0,\dots,\sigma_{k-1})^{T}  \left[e^{-i\frac{(\theta+\phi)}{2}\sigma_{k+1}} \bigl (c\mathbf{a}({\sigma_k})+is\sigma_{k+1}\mathbf{a}({-\sigma_k})\bigr )\right] \ket{\vec{\sigma}}.
\end{align}
The factor in the square brackets in the final line results from the action of the $k$'th block $C_{k,k+1}^{\theta}R^{\theta+\phi}_{k+1}$, and can be expressed compactly as 
\begin{equation}
    e^{-i\frac{(\theta+\phi)}{2}\sigma_{k+1}} \Bigl (c\mathbf{a}({\sigma_k})+is\sigma_{k+1}\mathbf{a}({-\sigma_k})\Bigr )=\begin{bmatrix}\mathbf{a}({\sigma_k}) &\mathbf{a}({-\sigma_k})\end{bmatrix}\begin{bmatrix}ce^{-i\frac{(\theta+\phi)}{2}\sigma_{k+1}} \\ is\sigma_{k+1}e^{-i\frac{(\theta+\phi)}{2}\sigma_{k+1}}\end{bmatrix}\equiv A^{\sigma_k}\mathbf{a}({\sigma_{k+1}})\label{eq:supp_Ak_def}
%    &= \begin{bmatrix}ce^{-i\frac{(\theta+\phi)}{2}\sigma_k}& ce^{\frac{i(\theta+\phi)}{2}\sigma_k}\\  is\sigma_ke^{-i\frac{(\theta+\phi)}{2}\sigma_k}&-is\sigma_ke^{\frac{i(\theta+\phi)}{2}\sigma_k}\end{bmatrix}\begin{bmatrix}ce^{-i\frac{(\theta+\phi)}{2}\sigma_{k+1}}\\is\sigma_{k+1}e^{-i\frac{(\theta+\phi)}{2}\sigma_{k+1}}\end{bmatrix}, \\
%    &\equiv A^{\sigma_k}\mathbf{a}({\sigma_{k+1}}).\label{eq:supp_Ak_def}
\end{equation} 
where we identified the vector $\mathbf{a}({\sigma_{k+1}})$ from Eq.~\eqref{eq:supp_ba_def} and introduced the matrix
\begin{align}
	A^{\sigma_k} &= \begin{bmatrix}\mathbf{a}({\sigma_k})& \mathbf{a}({-\sigma_k})\end{bmatrix}.
	%A^+&= \begin{bmatrix}c & c\\is & -is\end{bmatrix}\begin{bmatrix}e^{-i(\theta+\phi)/2} & 0\\ 0& e^{i(\theta+\phi)/2}\end{bmatrix},\label{eq:Apy}\\
	%A^-&= \begin{bmatrix}c & c\\-is & is\end{bmatrix}\begin{bmatrix}e^{i(\theta+\phi)/2} & 0\\ 0& e^{-i(\theta+\phi)/2}\end{bmatrix}.\label{eq:Amy}
\end{align}
Explicitly, this becomes
\begin{align}
	A^+&= \begin{bmatrix}c & c\\is & -is\end{bmatrix}\begin{bmatrix}e^{-i(\theta+\phi)/2} & 0\\ 0& e^{i(\theta+\phi)/2}\end{bmatrix},\label{eq:Apy}\\
	A^-&= \begin{bmatrix}c & c\\-is & is\end{bmatrix}\begin{bmatrix}e^{i(\theta+\phi)/2} & 0\\ 0& e^{-i(\theta+\phi)/2}\end{bmatrix}.\label{eq:Amy}
\end{align}
These are the matrices $A^+$ and $A^-$ from Eq.~(5) in the main text. Taken together, we have shown that
\begin{align}
    \ket{\psi_o} &=\nf  R_0^{-\phi}U(k)\sum_{\vec{\sigma}} \mathbf{r}(\sigma_0,\dots,\sigma_{k-1})^{T} \mathbf{a}({\sigma_k})  \ket{\vec{\sigma}}\\
     &=\nf R_0^{-\phi}U(k+1)\sum_{\vec{\sigma}}   \mathbf{r}(\sigma_0,\dots,\sigma_{k-1})^{T} A^{\sigma_k}\mathbf{a}({\sigma_{k+1}}) \ket{\vec{\sigma}}\\
     &\equiv\nf R_0^{-\phi}U(k+1)\sum_{\vec{\sigma}}   \mathbf{r}(\sigma_0,\dots,\sigma_{k})^{T}\mathbf{a}({\sigma_{k+1}}) \ket{\vec{\sigma}}.\label{eq:finalequivalence}
\end{align}
This establishes the validity of the general form of $\ket{\psi_o}$ in Eq.~\eqref{eq:psi0intermsofr} and yields the recurrence relation
\begin{equation}
    \mathbf{r}(\sigma_0,\dots,\sigma_{k})^{T} = \mathbf{r}(\sigma_0,\dots,\sigma_{k-1})^{T} A^{\sigma_k}\qquad\text{with}\qquad\mathbf{r}(\sigma_0)^{T} = \mathbf{b}(\sigma_0)^{T},
\end{equation}
which can be solved to produce
\begin{equation}
	\mathbf{r}(\sigma_0,\dots,\sigma_{k})^{T}=\mathbf{b}(\sigma_0)^{T}A^{\sigma_1}A^{\sigma_2}\cdots A^{\sigma_{k}}.
\end{equation}
Choosing $k=N-1$ in Eq.~\eqref{eq:finalequivalence} and using $U(N)=I$ and $\sigma_N=\sigma_0$ yields
\begin{equation}
    \ket{\psi_o}=\nf\sum_{\vec{\sigma}} \mathbf{b}(\sigma_0)^{T}A^{\sigma_1}A^{\sigma_2}\cdots A^{\sigma_{N-1}}(e^{i\frac{\phi}{2}\sigma_0}\mathbf{a}({\sigma_{0}})) \ket{\vec{\sigma}}.
\end{equation}
Finally, we express the expansion coefficients as traces, which leads to 
%Noticing that the trace of a scalar is the scalar itself, we can write the state as
\begin{align}
    \ket{\psi_o} &=\nf\sum_{\vec{\sigma}} \Tr\Bigl(B^{\sigma_0}A^{\sigma_1}A^{\sigma_2}\cdots A^{\sigma_{N-1}}\Bigr) \ket{\vec{\sigma}},\label{eq:mpsansatzy}
\end{align}
where 
\begin{align}
    B^{\sigma_0}&\equiv e^{i\frac{\phi}{2}}\mathbf{a}({\sigma_{0}})\mathbf{b}(\sigma_{0})^{T}.
\end{align}
Written out, we have
\begin{align}
	%B^{\sigma_0}&\equiv e^{i\frac{\phi}{2}}\mathbf{a}({\sigma_{0}})\mathbf{b}(\sigma_{0})^{T},\\
	B^+&=\begin{bmatrix}c & c\\ is & is\end{bmatrix}\begin{bmatrix}e^{-i(\theta+\phi)/2} & 0\\ 0& e^{-i(\theta-\phi)/2}\end{bmatrix}, \label{eq:Bpy}\\
	B^-&= \begin{bmatrix}c & c\\ -is & -is\end{bmatrix}\begin{bmatrix}e^{i(\theta+\phi)/2} & 0\\ 0& e^{i(\theta-\phi)/2}\end{bmatrix}.\label{eq:Bmy}
\end{align}
These are the matrices $B^+$ and $B^-$ from Eqs.~$(6)$ and $(7)$ in the main text.

\section{Derivation of expectation values}
Here we provide details of the derivation of analytic expressions for various expectation values appearing in the main text. 

\subsection{Trace formulas for expectation values}
We first demonstrate how the expectation values of one- and two-site observables with respect to a matrix product state of the form
\begin{equation}
	\ket{\psi_t}=\frac{1}{M}\sum_{\vec{\sigma}}\Tr(A^{\sigma_0}A^{\sigma_1}\cdots A^{\sigma_{N-1}})\ket{\vec{\sigma}\,}
\end{equation}
can be expressed as traces of transfer matrix products. Here $\ket{\vec{\sigma}\,}=\ket{\sigma_0,\sigma_1,\ldots,\sigma_{N-1}}$ with $\sigma_i\in\{+,-\}$ is a product state for $N$ spin sites, $A^\pm$ are matrices, and $M^{-1}$ is a normalisation factor. Let $O$ be a single-site observable, acting on the state $\ket{\sigma}$ of the site as
\begin{equation}
	%O_i\ket{\sigma_i}=\mel{+}{O}{\sigma_i}\ket{+}_i+\mel{-}{O}{\sigma_i}\ket{-}_i.
	O\ket{\sigma}=\mel{+}{O}{\sigma}\ket{+}+\mel{-}{O}{\sigma}\ket{-}.
\end{equation}
If $O_i$ denotes the observable $O$ for site $i$, then its action on $\ket{\psi_t}$ is
\begin{equation}
	O_i\ket{\psi_t}=\frac{1}{M}\sum_{\vec{\sigma}}\Tr(A^{\sigma_0}A^{\sigma_1}\cdots A^{\sigma_{N-1}})\left[\mel{+}{O}{\sigma_i}\ket{+}_i+\mel{-}{O}{\sigma_i}\ket{-}_i\right]\otimes\ket{\vec{\sigma}\,'},
\end{equation}
where $\ket{\vec{\sigma}\,'}$ excludes the state of site $i$. Taking the inner product with $\ket{\psi_t}$ then produces
\begin{equation}
	\mel{\psi_t}{O_i}{\psi_t}=\frac{1}{M^2}\sum_{\vec{\sigma}}\Tr\left(A^{\sigma_0}\cdots A^{\sigma_{i-1}}%
	\left[\mel{+}{O}{\sigma_i}^*A^++\mel{-}{O}{\sigma_i}^*A^-\right]%
	A^{\sigma_{i+1}}\cdots A^{\sigma_{N-1}}\right)^*\Tr(A^{\sigma_0}A^{\sigma_1}\cdots A^{\sigma_{N-1}}).	
	%\frac{1}{M^2}\sum_{\vec{\sigma}}\left[\mel{+}{O}{\sigma_i}\ket{+}_i+\mel{-}{O}{\sigma_i}\ket{-}_i\right]\Tr(A^{\sigma_0}A^{\sigma_1}\cdots A^{\sigma_{N-1}})
\end{equation}
From here we use the fact that for matrices $Q$ and $R$ it holds that $\Tr(Q\otimes R)=\Tr(Q)\Tr(R)$ to combine the product of the two traces in the sum above, and then $(Q_1Q_2)\otimes(R_1R_2)=(Q_1\otimes R_1)(Q_2\otimes R_2)$ to group matrices associated with the same site together in Kronecker products. The sum over $\vec{\sigma}$ now factorises within the trace, and can be performed to produce
\begin{equation}
	\mel{\psi_t}{O_i}{\psi_t}=\frac{1}{M^2}\Tr(T^i T_o T^{N-i-1}),\label{eq:1bodyEVgeneral}
\end{equation}
with the two transfer matrices given by
\begin{equation}
	T=A^{+*}\otimes A^++A^{-*}\otimes A^- \qquad\text{and}\qquad T_o=\sum_{\sigma,\sigma'=\pm}\mel{\sigma'}{O}{\sigma}A^{\sigma'*}\otimes A^\sigma.
\end{equation}
This result readily generalises to arbitrary multi-site observables. For example, 
\begin{equation}
	\mel{\psi_t}{O_i O_{i+r}}{\psi_t}=\frac{1}{M^2}\Tr(T^i T_o T^{r-1} T_o T^{N-i-r-1}).\label{eq:2bodyEVgeneral}
\end{equation}

\subsection{Transfer matrices - definitions and diagonalisation}
We will use the following compact notation for trigonometric functions of the two variational angles $\theta$ and $\phi$ that appear in our ansatz:
\begin{equation}\label{eq:cdsteu}
    \begin{aligned}
    c &\equiv \cos\bigl(\tfrac{\theta}{2}\bigr)  &  s &\equiv \sin\bigl(\tfrac{\theta}{2}\bigr)& C &\equiv \cos\bigl(\theta\bigr) & S &\equiv \sin\bigl(\theta\bigr)\\
    d &\equiv \cos\bigl(\tfrac{\theta+\phi}{2}\bigr) \qquad &   t &\equiv \sin\bigl(\tfrac{\theta+\phi}{2}\bigr) \qquad & D &\equiv \cos\bigl(\theta+\phi\bigr) \qquad &T &\equiv \sin\bigl(\theta+\phi\bigr)\\
    e &\equiv \cos\bigl(\tfrac{\theta-\phi}{2}\bigr)  &   u &\equiv \sin\bigl(\tfrac{\theta-\phi}{2}\bigr) & E &\equiv \cos\bigl(\theta-\phi\bigr)  & U &\equiv \sin\bigl(\theta-\phi\bigr)\\
    f &\equiv \cos\bigl(\tfrac{\phi}{2}\bigr)  &   v &\equiv \sin\bigl(\tfrac{\phi}{2}\bigr) & F &\equiv \cos\bigl(\phi\bigr)  &	V &\equiv \sin\bigl(\phi\bigr)
    \end{aligned}
\end{equation}
\emph{Note that this notation differs from the one used in the main text.} For calculating expectation values it will be convenient to switch to the $Z$ basis and rewrite the MPS form of the ansatz in Eq.~\eqref{eq:mpsansatzy} as a linear combination of $Z_i$ eigenstates. The simple connection between the eigenstates of $Y_i$ and $Z_i$ leads to the form
\begin{align}
	\ket{\psi_o}&=\sum_{\vec{\sigma}} \Tr(\Bz^{\sigma_0}\Az^{\sigma_1}\Az^{\sigma_2}\cdots \Az^{\sigma_{N-1}}) \ket{\vec{\sigma}\,},\label{eq:psi_o_mps}
\end{align} 
where $\ket{\sigma}\equiv\ket{z,\sigma}$, $\ket{\vec{\sigma}\,}=\ket{\sigma_0,\sigma_1,\sigma_2,\ldots}$ satisfies $Z_i\ket{\vec{\sigma}\,}=\sigma_i\ket{\vec{\sigma}\,}$ with $\sigma_i\in\{+,-\}$. Here $\Az^\pm$ and $\Bz^\pm$ are given in terms of the matrices $A^\pm$ and $B^\pm$ from Eqs.~\eqref{eq:Apy}, \eqref{eq:Amy}, \eqref{eq:Bpy} and \eqref{eq:Bmy} by
\begin{equation}
	\begin{bmatrix}A^+_z\\A^-_z\end{bmatrix}=\frac{1}{2}\begin{bmatrix}1 & 1\\i&-i\end{bmatrix}\begin{bmatrix}A^+\\A^-\end{bmatrix}, \qquad \begin{bmatrix}B^+_z\\B^-_z\end{bmatrix}=\frac{1}{2}\begin{bmatrix}1 & 1\\i&-i\end{bmatrix}\begin{bmatrix}B^+\\B^-\end{bmatrix}.
\end{equation}
Explicitly, we have 
\begin{equation}\label{eq:ABpm_matrices}
	\Az^+ = \begin{bmatrix} cd &  cd\\ st& st\end{bmatrix},\qquad \Az^- = \begin{bmatrix} ct &  -ct\\ -sd& sd\end{bmatrix},\qquad  \Bz^+ =  \begin{bmatrix} cd  &  ce \\  st & su \end{bmatrix},\qquad \Bz^- =\begin{bmatrix} ct  &  cu \\  -sd & -se \end{bmatrix}.
\end{equation} 
We also define the transfer matrix 
\begin{align}\label{eq:TaDefinition}
	T_a &= \Az^{+*}\otimes \Az^+ +\Az^{-*}\otimes \Az^-,
\end{align}
which is associated with the sites at positions $i>0$. Here $\Az^{+*}$ and $\Az^{-*}$ are the complex conjugates of $\Az^+$ and $\Az^-$. Similarly, the transfer matrix for the first site at $i=0$ reads
\begin{align}\label{eq:TbDefinition}
	T_b &= \Bz^{+*}\otimes \Bz^+ +\Bz^{-*}\otimes \Bz^-.
\end{align}
The transfer matrix associated with the observable $O_i$ for site $i>0$ is defined as
\begin{align}
	T_{ao} &\equiv\sum_{\sigma,\sigma^\prime =\pm} \Az^{\sigma*}\otimes \Az^{\sigma^\prime} \langle \sigma|O|\sigma^\prime\rangle,\label{eq:arbitrary_obs_transfera}
\end{align}
while for site $i=0$ this becomes
\begin{align}
	T_{bo} &\equiv\sum_{\sigma,\sigma^\prime =\pm} \Bz^{\sigma*}\otimes \Bz^{\sigma^\prime} \langle \sigma|O|\sigma^\prime\rangle.\label{eq:arbitrary_obs_transferb}
\end{align}
Our calculations will involve arbitrary powers of $T_a$ and $T_{ax}$, with the latter being the transfer matrix associated with the observable $X_i$ at a site $i>0$. It will therefore be useful to obtain diagonalised forms of these two matrices. From Eq.~\eqref{eq:ABpm_matrices} it is clear that the $\Az^\pm$ mattrices are at most of rank one, and since $\rank(M\otimes N)=\rank(M)\rank(N)$ and $\rank(M+ N)\leq\rank(M)+\rank(N)$ it follows from  Eq.~\eqref{eq:TaDefinition} that $T_a$ is at most of rank two. The two right and two left eigenvectors of $T_a$ with non-zero eigenvalues are found to be
\begin{equation}
	\ket{r_0}=(c^2,0,0,s^2)^T,\qquad\ket{r_1}=\left(\frac{c^2 D (C+S T-1)}{S T (S T-1)},\frac{1}{2},\frac{1}{2},\frac{s^2D (C-S T+1)}{S T (S T-1)}\right)^T\label{eq:r0r1}
\end{equation}
and
\begin{equation}
	\ket{l_0}=\left(1,\frac{C D}{1-S T},\frac{C D}{1-ST},1\right)^T,\qquad\ket{l_1}=\left(0,1,1,0\right)^T.
\end{equation}
These satisfy the eigenvalue equations
\begin{equation}
	T_a\ket{r_0}=\ket{r_0}, \qquad T_a\ket{r_1}=ST\ket{r_1},  \qquad  \bra{l_0}T_a=\bra{l_0},  \qquad  \bra{l_1}T_a=ST\bra{l_1}, 
\end{equation}
and are normalised such that $\bra{l_i}\ket{r_j}=\delta_{ij}$ for $i,j\in\{0,1\}$. This allows $T_a$, and powers thereof, to be expressed in the diagonal forms 
\begin{equation}\label{eq:Ta_eig}
	T_a=\dyad{r_0}{l_0} + ST\,\dyad{r_1}{l_1}\quad\text{and}\quad T_a^{\,N}=\dyad{r_0}{l_0} + (ST)^{N}\dyad{r_1}{l_1}.
\end{equation}
The same procedure can be applied to the transfer matrix $T_{ax}$, of which the explicit form follows from Eq.~\eqref{eq:arbitrary_obs_transfera} as 
\begin{align}
	T_{ax}&\equiv \Az^{+*}\otimes \Az^- + \Az^{-*}\otimes \Az^+.\label{eq:Tx_eig}
\end{align}
Again $T_{ax}$ is at most of rank two, and the eigenvectors with non-zero eigenvalues are
\begin{equation}
	\ket{{r^{\prime}_0}}=(0,-1,1,0)^T,\qquad\ket{{r^{\prime}_1}}=\left(-c^2,\frac{DS}{2T},\frac{DS}{2T},s^2\right)^T
\end{equation}
and
\begin{equation}
	\ket*{{l^{\prime}_0}}=\left(0,-\frac{1}{2},-\frac{1}{2},0\right)^T,\qquad\ket*{{l^{\prime}_1}}=\left(-1,0,0,1\right)^T.
\end{equation}
These obey the eigenvalue equations
\begin{equation}
	T_{ax}\ket{{r^{\prime}_0}}=S\ket{{r^{\prime}_0}}, \qquad T_{ax}\ket{{r^{\prime}_1}}=T\ket{{r^{\prime}_1}},  \qquad  \bra*{{l^{\prime}_0}}T_{ax}=S\bra*{{l^{\prime}_0}},  \qquad  \bra*{{l^{\prime}_1}}T_{ax}=T\bra*{{l^{\prime}_1}}, 
\end{equation}
and are normalised such that $\bra*{{l^{\prime}_i}}\ket*{{r^{\prime}_j}}=\delta_{ij}$ for $i,j\in\{0,1\}$. This allows for the diagonal representations
\begin{equation}\label{eq:Tax_eig}
	T_{ax}=S\dyad*{{r^{\prime}_0}}{{l^{\prime}_0}} + T\,\dyad*{{r^{\prime}_1}}{{l^{\prime}_1}}\quad\text{and}\quad T_{ax}^{\,N}=S^N\dyad*{{r^{\prime}_0}}{{l^{\prime}_0}} + T^{N}\dyad*{{r^{\prime}_1}}{{l^{\prime}_1}}.
\end{equation}
\subsection{Expectation value calculations for the translationally invariant ansatz}
Recall that the translationally invariant ansatz is obtained by replacing $B^{\pm}$ by $A^{\pm}$ in the original MPS ansatz \eqref{eq:mpsansatzy}. This yields
\begin{align}
    \ket*{\widetilde{\psi}_t}&\equiv \sum_{\vec{\sigma}} \Tr(\Az^{\sigma_0}\cdots \Az^{\sigma_{N-1}})\ket{\vec{\sigma}\,}, \label{eq:psi_t_mp_unnormalised}
\end{align}
where the tilde indicates that the state is unnormalised. The loss of normalisation is due to the fact that this state can no longer be expressed as a unitary operator acting on a product state, as was the case for the original ansatz in Eq.~\eqref{eq:supp_anzcircuit}. To normalise the state in Eq.~\eqref{eq:psi_t_mp_unnormalised} we first compute, using Eq.~\eqref{eq:1bodyEVgeneral} with $O=I$, 
\begin{equation}
    M_t^2 \equiv \braket*{\widetilde{\psi}_t}{\widetilde{\psi}_t}=\sum_{\vec{\sigma}} \Tr(\Az^{\sigma_0}\cdots \Az^{\sigma_{N-1}})^*\Tr(\Az^{\sigma_0}\cdots \Az^{\sigma_{N-1}})=\Tr(T_a^{N}).\label{eq:traceTaN}
\end{equation}
Using the diagonal form of $T_a$ in Eq.~\eqref{eq:Ta_eig} we obtain
\begin{equation}
    M_t^2 = 1+ (ST)^N,\label{eq:translational_norm}
\end{equation}
which we use to define the normalised translationally invariant state as 
\begin{equation}
    \ket{\psi_t} = \frac{1}{M_t}\ket*{\widetilde{\psi}_t}.\label{eq:psi_t_mps}
\end{equation}
The expectation value of the observable $X$ at site $i$ with respect to $\ket{\psi_t}$ now follows from Eq.~\eqref{eq:1bodyEVgeneral} as 
\begin{equation}
    \expval{X_i} = \frac{1}{M_t^2}\Tr(T_a^{i}T_{ax}T_a^{N-i-1}).
\end{equation}
The explicit forms of $T_a$, $T_{ax}$, and their powers in Eqs.~\eqref{eq:Ta_eig} and \eqref{eq:Tax_eig} allow this and similar traces to be evaluated easily. The resulting algebra, while somewhat lengthy in cases, is straightforward to perform using software packages such as \emph{Mathematica} or \emph{SymPy}. We find that
\begin{equation}
	\expval{X_i}=\frac{1}{M_t^2}\left[\frac{C^2(S-T)}{ST-1}+(ST)^N\frac{D^2(S-T)}{T^2(ST-1)}\right],\label{eq:trans_xexpval}
\end{equation}
which corresponds to Eq.~(9) in the main text.\\

We can apply the same procedure to obtain a closed-form expression for the correlation function $\expval{\psi_t|Z_iZ_{i+r}|\psi_t}$, which, from Eq.~\eqref{eq:1bodyEVgeneral} is given by the trace
\begin{equation}
    \expval{\psi_t|Z_iZ_{i+r}|\psi_t} = \frac{1}{M_t^2}\Tr\bigl(T_{az}T_a^{r-1}T_{az}T_a^{N-r-1}\bigr).\label{eq:expval_zzr_trans}
\end{equation}
The relevant transfer matrix follows from Eq.~\eqref{eq:arbitrary_obs_transfera} as
\begin{equation}
    T_{az} \equiv \Az^{+*}\otimes \Az^+ - \Az^{-*}\otimes \Az^-.
\end{equation}
Combining this with the diagonal form of $T_a$ from Eq.~\eqref{eq:Ta_eig} yields
\begin{align}
	\expval{\psi_t|Z_iZ_{i+r}|\psi_t} &=\frac{1}{M^2_t}\left[f(r)+ (ST)^Nf(-r)\right],
\end{align}
where
\begin{align}\label{eq:fdefinition}
	f(r) &= \frac{C^2D^2+(S-T)^2(ST)^r}{(ST-1)^2}.
\end{align}
These are Eqs.~(11) and~(12) in the main text.

\subsection{Expectation value calculations for the translationally and parity symmetric ansatz}
We can restore the parity symmetry that the state $\ket*{\widetilde{\psi}_t}$ from Eq.~\eqref{eq:psi_t_mps} breaks by projecting it onto the positive parity subspace. This yields the unnormalised state
\begin{equation}
\ket*{\widetilde{\psi}_p}\equiv\frac{1}{\sqrt{2}}\Bigl(\ket*{\widetilde{\psi}_t} +P\ket*{\widetilde{\psi}_t}\Bigr),
\end{equation}
where $P\equiv\prod_{i=0}^{N-1} X_i$ is the parity operator that flips all spins via the Pauli $X$ operator. The required normalisation factor for $\ket*{\widetilde{\psi}_p}$ is found to be 
\begin{equation}
M_p^2\equiv \braket*{\widetilde{\psi}_p}{\widetilde{\psi}_p}=\braket*{\widetilde{\psi}_t}{\widetilde{\psi}_t}+\expval*{\widetilde{\psi}_t|P|\widetilde{\psi}_t}=M_t^2+\Tr(T_{ax}^N),
\end{equation}
where the extension of Eq.~\eqref{eq:2bodyEVgeneral} to the $N$-site operator $P$ was used to express the second term as a trace. Using the normalisation factor of $\ket*{\psi_t}$ in Eq.~\eqref{eq:translational_norm} and the diagonal form of $T_{ax}$ in Eq.~\eqref{eq:Tx_eig} we find that
\begin{equation}
 M_p^2= 1+(ST)^N+T^N+S^N.
\end{equation}
The normalised parity and translationally symmetric ansatz is now given by
\begin{equation}
    \ket{\psi_p} = \frac{1}{M_p}\ket*{\widetilde{\psi}_p}. \label{eq:psi_p_mps}
\end{equation}
To calculate the expectation value of the observable $X_i$ with respect to this state, we first write
\begin{align}
    \expval*{\psi_p|X_i|\psi_p}&=\frac{1}{M_p^2}\Bigl(\expval*{\widetilde{\psi}_t|X_i|{\widetilde{\psi}}_t}+\expval*{\widetilde{\psi}_t|X_iP|\widetilde{\psi}_t}\Bigr). \label{eq:expval_x_parity_sub}
\end{align}
As in the previous section, we can translate these expectation values with respect to $\ket*{\widetilde{\psi}_t}$ into traces involving the relevant transfer matrices. Notice that $X_i$ in the second term undoes the action of $P$ on the $i^{\rm th}$ site, and so by extending  Eq.~\eqref{eq:2bodyEVgeneral},  Eq.~\eqref{eq:expval_x_parity_sub} becomes
\begin{align}
    \expval{\psi_p|X_i|\psi_p}&=\frac{1}{M_p^2}\left[\Tr\bigl(T_a^{N-1}T_{ax}\bigr) + \Tr\bigl(T_{ax}^{N-1}T_a\bigr) \right].
\end{align}
The diagonal forms of $T_a$ and $T_{ax}$ in Eqs.~\eqref{eq:Ta_eig} and \eqref{eq:Tax_eig} now lead to 
\begin{equation}
	\expval{\psi_p|X_i|\psi_p}=\frac{1}{M_p^2}\left[\frac{C^2 (S-T)}{S T-1}+\frac{(S T)^N D^2 (S-T)}{T^2 (S T-1)}+T^{N-2} (T-S) (S T+1)\right].\label{eq:expval_x_parity}
\end{equation}
The same procedure yields the correlation function
\begin{equation}
    \expval{\psi_p|Z_i Z_{i + r}|\psi_p} = \frac{1}{M^2_p} \left[ f(r) + (ST)^N f(-r) \right] + \frac{(ST)^r}{M^2_p} \left( S^{N - 2r} + T^{N - 2r} \right),
\end{equation}
with $f(r)$ as in Eq.~\eqref{eq:fdefinition}.
\subsection{Expectation value calculations for the original ansatz}
The expression $\mel{\psi_t}{O_i}{\psi_t}=\frac{1}{M^2}\Tr(T^i T_O T^{N-i-1})$ in Eq.~\eqref{eq:1bodyEVgeneral} for the expectation value of a single-site operator with respect to a translationally invariant MPS state can be easily adapted to the original ansatz $\ket{\psi_o}$ in Eq.~\eqref{eq:mpsansatzy}. For the expectation value of $X_i$ this results in
\begin{align}
    \expval{\psi_o|X_i|\psi_o}
    &=
    \begin{cases}
      \Tr\bigl(T_b T_a^{i-1} T_{x} T_a^{N-i-1} \bigr) & \text{when } i \neq 0, \\[4pt]
      \Tr\bigl(T_{bx} T_a^{N-1}\bigr)                  & \text{when } i = 0.
    \end{cases} \label{eq:expval_x_original}
\end{align}
The transfer matrix $T_b$ is given in Eq.~\eqref{eq:TbDefinition}, while $T_{bx}$ follows from Eq.~\eqref{eq:arbitrary_obs_transferb} as
\begin{equation}
	T_{bx} = B_z^{+*}\otimes \Bz^- + \Bz^{-*}\otimes \Bz^+.
\end{equation}
For $i\neq 0$ we find 
\begin{align}
	\expval{\psi_o|X_i|\psi_o}&= \frac{C^2(S-T)}{ST-1}-(ST)^{i-1}\Bigl(\frac{C^2D^2S}{ST-1}+SCDF\Bigr).\label{eq:expval_x_orig}
\end{align}
For $i=0$ this becomes
\begin{align}
	\expval{\psi_o|X_0|\psi_o}&= C\!\left(c^{2} T + s^{2} U\right)
	+ \frac{C S D}{S T - 1}\!\left(c^{2} D + s^{2} E\right) -CS(S T)^{N-1}\\[4pt]
	&\quad
	-(S T)^{N-1}\left[\frac{c^2D^2(C+ST-1)+s^2DE(C-ST+1)}{T(ST-1)}\right].
\end{align}
\subsection{Thermodynamic limit of the transverse field expectation value}
The expectation values of $X_i$ with respect to the three main ansatzes were found in Eqs.~\eqref{eq:trans_xexpval}, \eqref{eq:expval_x_parity} and \eqref{eq:expval_x_orig} to be
\begin{align}
    \expval{\psi_t|X_i|\psi_t} &=\frac{1}{M_t^2}\left[\frac{C^2(S-T)}{ST-1}+(ST)^N\frac{D^2(S-T)}{T^2(ST-1)}\right],\label{eq:tran_xexpval2}\\
    \expval{\psi_p|X_i|\psi_p}&=\frac{1}{M_p^2}\left[\frac{C^2(S-T)}{ST-1} +T^{N}\frac{(D^2S^N+1-S^2T^2)(S-T)}{T^2(ST-1)}\right],\label{eq:expval_x_parity2}\\
    \expval{\psi_o|X_{i\neq0}|\psi_o}&=\frac{C^2(S-T)}{ST-1}-(ST)^{i-1}\Bigl(\frac{C^2D^2S}{ST-1}+SCDF\Bigr).\label{eq:expval_x_orig2}
\end{align}
We see that both the translational and parity symmetric ansatzes converge to the same expression in the thermodynamic limit:
\begin{align}
\lim_{N\rightarrow\infty}  \expval{\psi_t|X_i|\psi_t} = \lim_{N\rightarrow\infty}  \expval{\psi_p|X_i|\psi_p} &= \frac{C^2(S-T)}{ST-1}.
\end{align}
In particular, this implies that the average magnetisation per spin in the direction of the transverse field, i.e. $\expval{\sigma_x}$ with $\sigma_x=\frac{1}{2}\sum_{i=0}^{N-1}X_i$, will also be given by the same expression in the thermodynamic limit for both these ansatzes. This term appears in the Hamiltonian for both the Ising and LMG models. If we consider the same quantity with respect to the original ansatz we find that it too converges to this expression:
\begin{align}
    \lim_{N\rightarrow\infty}\frac{1}{N}\expval{\psi_o|\sigma_x|\psi_o}&= \lim_{N\rightarrow\infty}\Bigl(\frac{1}{2N}\expval{\psi_o|X_0|\psi_o}+\frac{1}{2N}\sum_{i=1}^{N-1}\expval{\psi_o|X_i|\psi_o}\Bigr),\\
    &=\frac{C^2(S-T)}{2(ST-1)} -\lim_{N\rightarrow\infty}\frac{K_0}{2N}\sum_{i=1}^{N-1}(ST)^{i-1},\\
    &=\frac{C^2(S-T)}{2(ST-1)} -\lim_{N\rightarrow\infty}\frac{K_0}{2N}\frac{1-(ST)^{N-1}}{1-ST},\\
    &=\frac{C^2(S-T)}{2(ST-1)}.
\end{align}
Here $K_0= \frac{C^2D^2S}{ST-1}+SCDF$ which does not depend on $N$.
\section{Further details on calculations for the LMG model}
\subsection{The energy per spin in the thermodynamic limit}
Both $S=\sin(\theta)$ and $T=\cos(\theta+\phi)$ are sines of angles, and so $ST$ cannot exceed one in magnitude. Starting from the full expression for the expectation value of the LMG Hamiltonian and assuming that $|ST|<1$ then yields in the $N\rightarrow\infty$ limit the result
\begin{equation}
	\lim_{N\rightarrow\infty}\ave{H}/N=\frac{h \left(S^2-1\right) (S-T)}{2 S T-2}-\frac{J \left(S^2-1\right) \left(T^2-1\right)}{8 (S T-1)^2},
\end{equation}
as given in Eq.~$(17)$ of the paper. The case for which $|ST|=1$ occurs when $(S,T)=(1,1)$, $(S,T)=(-1,-1)$, $(S,T)=(1,-1)$, or $(S,T)=(-1,1)$. Focussing on even $N$, we find for the first two of these that $\lim_{N\rightarrow\infty}\ave{H}/N=-J/8$, and for the second two $\lim_{N\rightarrow\infty}\ave{H}/N=0$. It is only precisely at $h=0$ where the first of these energies match the energy obtained from the optimal angles in Eq.~ (18) of the main text. For all other values of $h$, using values of the angles for which $|ST|=1$ therefore does not yield the minimum energy. Throughout the main text we therefore assume that $|ST|<1$.

\subsection{Energy minimisation}
For the LMG model the derivatives of the energy expectation value with respect to the two angles, written in terms of $S=\sin(\theta)$ and $T=\sin(\theta+\phi)$ are:
\begin{align}
	\partial_\theta \ave{H}/N&=\frac{\sqrt{1-S^2} \left(2 h (S T-1) \left(-\left(S^2+1\right) T^2+2 \left(S^2+1\right) S T-3 S^2+1\right)+\left(T^2-1\right) (S-T)\right)}{4 (S T-1)^3},\\
	\partial_\phi \ave{H}/N&=-\frac{\left(S^2-1\right) \sqrt{1-T^2} \left(2 h \left(S^2-1\right) (S T-1)+S-T\right)}{4(S T-1)^3}.
\end{align}
Setting both of these derivatives to zero yields the following solutions:
\begin{enumerate}
	\item $S=\sin(\theta)=\pm1$ with $T=\sin(\theta+\phi)$ arbitrary. This yields $\ave{H}/N=0$.
	\item $S=\sin(\theta)=0$ and $T=\sin(\theta+\phi)=\pm 1$. This yields $\ave{H}/N=\mp h/2$.
	\item $S=\sin(\theta)=0$ and $T=\sin(\theta+\phi)=2h$. This yields $\ave{H}/N=-(1+4h^2)/8$.
\end{enumerate}
The stationary point that yields the minimum value of $\ave{H}/N$ is therefore given by 
\begin{equation}
	\sin(\theta+\phi)=\begin{cases}
		2h& |2h|\leq1 \\
		{\rm sgn}(h)& \text{otherwise}
	\end{cases}\qquad\text{and}\qquad\sin(\theta)=0.
\end{equation}
This yields $\theta=0,\pi$. Selecting $\theta=0$ leads to one solution for $\phi$ when $|2h|\geq1$ and two solutions when $|2h|<1$. The latter two solutions correspond to the two symmetry-broken ground states in the ferromagnetic phase. Selecting $\theta=\pi$ leads to the same solution(s) for $\phi$, just shifted by $\pi$. However, this yields, up to a phase, the same ansatz as when $\theta=0$. (Although the two solutions for $\phi$ get exchanged.) In the main text we therefore only consider the $\theta=0$ solution.

\subsection{Symmetry Projection}
Here we provide details on how the permutation and parity symmetries of the LMG Hamiltonian is restored in the translationally invariant ansatz in Eq.~$(8)$ of the main text. This requires projecting the ansatz onto the appropriate symmetry subspace. We first define the permutation operator $\mathbb{P}$ through its action on a spin basis state $\ket{\vec{\sigma}\,}=\ket{\sigma_0,\sigma_1,\ldots,\sigma_{N-1}}$ as
\begin{equation}\label{eq:supp_Pdef}
	\mathbb{P}\ket{\vec{\sigma}\,}=\frac{1}{N!}\sum_{p\in \sigma_N}\ket{\sigma_{p(0)},\sigma_{p(1)},\ldots,\sigma_{p(N-1)}}
\end{equation}
where the sum runs over all permutations of $\{0,1,\ldots,N-1\}$. Let $n_-(\vec{\sigma})$ denote the number of $-$ states appearing in $\vec{\sigma}$. We define the normalised state
\begin{equation}
	\ket{n}=\binom{N}{n}^{-1/2}\sum_{\vec{\sigma}'\,}\delta_{n_-(\vec{\sigma}'),n}\ket{\vec{\sigma}'\,},
\end{equation}
which is an equal superposition of the product states with exactly $n$ spins in the $-$ state. This allows us to rewrite Eq.~\eqref{eq:supp_Pdef} as
\begin{equation}
 \mathbb{P}\ket{\vec{\sigma}\,}=\binom{N}{n}^{-1/2}\ket{n=n_-(\vec{\sigma})}.
\end{equation}
Applying $\mathbb{P}$ to the ansatz $\ket*{\psi_t}=M_t^{-1}\sum_{\vec{\sigma}} \Tr(\Az^{\sigma_0}\cdots \Az^{\sigma_{N-1}})\ket{\vec{\sigma}\,}$ produces
\begin{equation}\label{eq:supp_permutationsymmetric}
	\mathbb{P}\ket{\psi_t}=\frac{1}{M_t}\sum_{n=0}^{N}\binom{N}{n}^{-1/2}\Tr\left[\sum_{\vec{\sigma}\,}\delta_{n_-(\vec{\sigma}),n}\Az^{\sigma_0}\cdots \Az^{\sigma_{N-1}}\right]\ket{n},
\end{equation}
where the trace contains a sum of all products of $n$ and $(N-n)$ of the  $A_z^-$ and $A_z^+$ matrices, respectively. To calculate this trace we utilise a generating function approach. Specifically, we note that the trace appearing in Eq.~\eqref{eq:supp_permutationsymmetric} will be the coefficient of the $x^{n}$ term in the polynomial
\begin{align}
{\rm Tr}\bigl ( (A^+_z+xA^-_z)^N\bigr )&=\lambda_+(x)^N + \lambda_-(x)^N, \label{eq:gen_tr}
\end{align}
where $\lambda_\pm(x)$ are the two eigenvalues of $A^+_z+xA^-_z$. These can be calculated as
\begin{align}
    \lambda_{\pm}(x)&=\frac{1}{2}\left ( (a+xb) \pm \sqrt{(a+xb)^2 -4xg} \right ) \label{eq:gen_eigens}    
\end{align}
where
\begin{equation}
	a=\cos(\frac{\phi}{2}), \quad b=\sin(\theta +\frac{\phi}{2}), \quad g=\sin(\theta).
\end{equation}
We can expand \ref{eq:gen_tr} using the binomial theorem, which results in
\begin{align}
    \lambda_+^N+\lambda_-^N&=\frac{1}{2^{N-1}}\sum_{i=0}^{\lfloor N/2 \rfloor} \binom{N}{2i}\left [ (a+xb)^2-4xg \right ]^i(a+xb)^{N-2i}.
\end{align}
The goal now is to rewrite this sum in order to group like powers of $x$ together. Expanding the power of the square bracket in the sum yields
\begin{align}
    \lambda_+^N+\lambda_-^N&=\frac{1}{2^{N-1}}\sum_{i=0}^{\lfloor N/2 \rfloor} \binom{N}{2i}\sum_{j=0}^i \binom{i}{j} (-4g)^jx^j(a+xb)^{2(i-j)}(a+xb)^{N-2i}\\
                        &=\frac{1}{2^{N-1}}\sum_{i=0}^{\lfloor N/2 \rfloor} \binom{N}{2i}\sum_{j=0}^i\binom{i}{j} (-4g)^jx^j(a+xb)^{N-2j}.
\end{align}
Next we expand the $N-2j$ power of $(a+x b)$ to get
\begin{align}
    \lambda_+^N+\lambda_-^N&=\frac{1}{2^{N-1}}\sum_{i=0}^{\lfloor N/2 \rfloor} \binom{N}{2i}\sum_{j=0}^i\binom{i}{j} (-4g)^jx^j\sum_{k=0}^{N-2j}\binom{N-2j}{k}x^kb^ka^{N-2j-k}\\
    &=\frac{1}{2^{N-1}}\sum_{i=0}^{\lfloor N/2 \rfloor} \sum_{j=0}^i\sum_{k=0}^{N-2j}\binom{N}{2i}\binom{i}{j}\binom{N-2j}{k}a^{N-2j-k}b^k (-4g)^j x^{k+j}.
\end{align}
Finally, we replace the sum over $k\in\{0,\ldots,N-2j\}$ with one over $n=k+j\in\{j,\ldots,N-j\}$. This produces
\begin{align}
    \lambda_+^N+\lambda_-^N&=\frac{1}{2^{N-1}}\sum_{i=0}^{\lfloor N/2 \rfloor} \sum_{j=0}^i\sum_{n=j}^{N-j}\binom{N}{2i}\binom{i}{j}\binom{N-2j}{n-j}a^{N-j-n}b^{n-j} (-4g)^j x^{n}. \label{eq:before_ind}
\end{align}
To obtain an expression for the coefficient of $x^n$ we need to change the order of the summations. Here we use the fact that
\begin{align}
    \binom{m}{k}&=0 \quad\text{ whenever } m<k \text{ or } k<0.\label{eq:binom_zero}
\end{align}
From Eq.~\eqref{eq:binom_zero}  we see that $\binom{N-2j}{n-j}=0$ if $n<j$ so that 
\begin{align}
\sum_{n=j}^{N-j}\binom{N-2j}{n-j}&=\sum_{n=0}^{N-j}\binom{N-2j}{n-j},
\end{align}
where we chose $n=0\leq j$ as the lower limit of the summation. Similarly, the coefficient $\binom{N-2j}{n-j}$ is zero for $n-j>N-2j$, which is equivalent to $n>N-j$ so that
\begin{align}
\sum_{n=0}^{N-j}\binom{N-2j}{n-j}&=\sum_{n=0}^{N}\binom{N-2j}{n-j},
\end{align}
where we chose $n=N\geq N-j$ as the upper limit of the summation. Finally, we will show that
\begin{align}
    \sum_{j=0}^i \binom{i}{j}\binom{N-2j}{n-j}&=\sum_{j=0}^{n} \binom{i}{j}\binom{N-2j}{n-j},\label{eq:subshow}
\end{align}
where both $j=i$ or $j=n$ are valid upper limits for the summation. To see this, consider the case $i< n$, then for any $j=n> i$ we have $\binom{i}{j}=0$, making Eq.~\eqref{eq:subshow} true. Next, consider the case $n<i$, then for any $j=i>n$ we have $\binom{N-2j}{n-j}=0$ also making Eq.~\eqref{eq:subshow} true. Finally, Eq.~\eqref{eq:subshow} is obviously true when $i=n$. With these three observations we can rewrite Eq.~\eqref{eq:before_ind} as
\begin{align}
    \lambda_+^N+\lambda_-^N&=\frac{1}{2^{N-1}}\sum_{n=0}^{N}x^{n} a^{N-n}b^n \sum_{i=0}^{\lfloor N/2 \rfloor} \sum_{j=0}^n\binom{N}{2i}\binom{i}{j}\binom{N-2j}{n-j} (-1)^j\left ( \frac{4g}{ab} \right )^j \label{eq:sub1}
\end{align}
If we let
\begin{align}
    T(N,j)&\equiv\sum_{i=0}^{\lfloor N/2 \rfloor}\binom{N}{2i}\binom{i}{j},\label{eq:riordan_appendix}
\end{align}
then Eq.\eqref{eq:sub1} can be written as
\begin{align}
    \lambda_+^N+\lambda_-^N&=\sum_{n=0}^{N}x^{n} \frac{a^{N-n}b^n}{2^{N-1}}  \sum_{j=0}^n (-1)^jT(N,j)\binom{N-2j}{n-j}\left ( \frac{4g}{ab} \right )^j,\\
    & = \sum_{n=0}^{N}x^{n} S(N,n),
\end{align}
where
\begin{align}
    S(N,n)&\equiv \frac{a^{N-n}b^n}{2^{N-1}}  \sum_{j=0}^n (-1)^jT(N,j)\binom{N-2j}{n-j}\left ( \frac{4g}{ab} \right )^j \label{eq:swap_symm_amp_appendix}.
\end{align}
This results provides an explicit form for the amplitudes of the permutation symmetric projected ansatz 
\begin{equation}
	\mathbb{P}\ket{\psi_t}=\frac{1}{M_t}\sum_{n=0}^{N}\binom{N}{n}^{-1/2}S(N,n)\ket{n}.
\end{equation}
The parity transformation swaps $+$ and $-$ spin states, or equivalently, maps $\ket{n}$ to $\ket{N-n}$. This observation allows us to restore the parity symmetry in $\mathbb{P}\ket{\psi_t}$ above. The resulting (unnormalised) permutation and parity invariant ansatz reads
%This expression Eq.~\eqref{eq:swap_symm_amp_appendix} represents the amplitudes for an unnormalised swap symmetric state, restoring parity symmetry we find 
\begin{equation}
    \ket*{\widetilde{\psi}_s} = \frac{1}{M_t} \sum_{n=0}^N \binom{N}{n}^{-\frac{1}{2}} P(N,n) \ket{n}
\end{equation}
where
\begin{equation}
	P(N,n)\equiv \frac{1}{2} ( S(N,n)+S(N,N-n)).
\end{equation}
Interestingly, $T(N,j)$ in Eq.~\eqref{eq:riordan_appendix} turns out to be the Riordan array \cite{RiordanTriangle1x2024} which obeys the recurrence relation
\begin{align}
    T(N,j)=2T(N-1,j)+T(N-2,j-1).
\end{align}
Using this result alongside the binomial additive identity $\binom{N}{k}=\binom{N-1}{k}+\binom{N-1}{k-1}$ we can derive a recurrence relation for the $S(N,n)$ amplitudes as
\begin{align}
    S(N,n)&= \frac{a^{N-n}b^n}{2^{N-1}}  \sum_{j=0}^n (-1)^jT(N,j)\binom{N-2j}{n-j}\left ( \frac{4g}{ab} \right )^j\\
    &= \frac{a^{N-n}b^n}{2^{N-1}}  \sum_{j=0}^n \Bigl (2T(N-1,j)+T(N-2,j-1) \Bigr ) \left [ \binom{N-1-2j}{n-j}+\binom{N-1-2j}{n-1-j}\right ] (-1)^j\left ( \frac{4g}{ab} \right )^j\\
    &=aS(N-1,n)+bS(N-1,n-1)+\frac{a^{N-n}b^n}{2^{N-1}}\sum_{j=0}^n T(N-2,j-1) \binom{N-2j}{n-j} (-1)^j\left ( \frac{4g}{ab} \right )^j\\
    &= aS(N-1,n)+bS(N-1,n-1)-sS(N-2,n-1)
\end{align}
where $S(0,0)=2,S(1,0)=a,S(1,1)=b$ and $S(N,n)=0 \text{ if } N<0 \text{ or } n<0$ generates the triangle.

\section{Further details on calculations for the TFIM model}
For the Ising model the energy expectation value in the thermodynamic limit is given in Eq.~(A2) of the end matter of the main text. Its derivatives with respect to the two angles, written again in terms of $S=\sin(\theta)$ and $T=\sin(\theta+\phi)$ are:
\begin{align}
	\partial_\theta \ave{H}/N&=\frac{\sqrt{1-S^2} \left(h \left(-2 \left(S^2+1\right) T^2+4\left(S^2+1\right) ST-6 S^2+2\right)-\left(S^2+2\right) T+2 S+T^3\right)}{4 (ST-1)^2},\\
	\partial_\phi \ave{H}/N&=\frac{\sqrt{1-T^2} \left((S-T) (S (S+T)-2)-2 h \left(S^2-1\right)^2\right)}{4 (ST-1)^2}.
\end{align}

Setting these to zero leads to the following solutions for the stationary points:
\begin{enumerate}
	\item For $h\in\mathbb{R}$: $(S,T)=(1,1)$ and $(-1,-1)$ yielding $\ave{H}/N=-1/4$
	\item For $h\in\mathbb{R}$: $(S,T)=(1,-1)$ and $(-1,1)$ yielding $\ave{H}/N=1/4$
	\item For $|h|\geq 1/4$: $(S,T)=(1/(4h),1)$ yielding $\ave{H}/N=-h/2-1/(32 h)$
	\item For $|h|\geq 1/4$: $(S,T)=(1/(4h),-1)$ yielding $\ave{H}/N=h/2+1/(32 h)$
\end{enumerate}
The remaining solution satisfies 
\begin{equation}
	h(1+S^2)^2-2S=0 \quad\text{and}\quad T=S(4/(1+S^2)-1).
\end{equation}
Since $|T|\leq 1$, the second equation places the restriction $|S|\leq \sqrt{2}-1$ on $S$. In turn, this limits, via the first equation, the range of $h$ values for which this solution exists to $|h|\leq h_c=(\sqrt{2}+1)/4$. The first equation is a quartic polynomial in $S$, and yields four solutions. Only one of these is real and satisfies $|S|\leq \sqrt{2}-1$. We denote this solution by $S^*(h)$ and the corresponding solution for $T$ by $T^*(h)$. Comparing the values of $\ave{H}/N$ at these stationary points allows us to identify the values of $S$ and $T$ that minimise $H$:
\begin{equation}
	S=\begin{cases}
		1/(4h) & |h|>h_c \\
		S^*(h) & |h|\leq h_c \\
	\end{cases}\qquad\text{and}\qquad T=\begin{cases}
		{\rm sgn}(h) & |h|>h_c \\
		T^*(h) & |h|\leq h_c \\
	\end{cases}
\end{equation}
When $|h|\geq h_c$ we therefore have $|T|=1$, which implies that $D=0$, since $T^2+D^2=1$. This causes the expectation value $\ave{Z_i}$ in Eq.~(9) of the main text to vanish.

\section{Domain-Specific Language (DSL) Examples}
The DSL used in the main text is implemented as an open-source Python package \cite{lourensHierarqcalGithub2024}, which was first presented in \cite{lourensHierarchicalQuantumCircuit2023}. Ref.~\cite{lourensHierarchicalQuantumCircuit2023} introduced the concept of abstract quantum circuit structures that can be composed to form new structures. In this work, we extend that abstraction to include arbitrary tensor networks and develop a search algorithm to obtain analytically tractable ansatzes for quantum many-body states. From a technical perspective, the package was updated accordingly to support this extension. Below, we provide examples of the DSL usage.\\\\
From the package, we will make use of the following classes and function:
\begin{lstlisting}[style=pyclasses]
from hierarqcal import Qcycle, Qpivot, Qmask, Qinit, plot_circuit
\end{lstlisting}
In the main text we mention that primitives are defined by size-independent properties, the main two being an edge generation pattern with an associated tensor. In Figs.~\ref{fig:cycle-n4} and~\ref{fig:cycle-n5} we show an example of a cycle primitive, named \emph{motif\_1}. The associated tensor is $e^{-i\frac{\theta}{4}Z\otimes Y}$ which is rank $4$ and specified with the \emph{mapping} argument. The edge generation pattern is specified through the arguments for \emph{Qcycle}, namely: \emph{stride} and \emph{step} (the other arguments take on default values). This results in a pattern that connects to every second site, starting from the first site, moving from one site to the next. The  motif, which is independent of system size, can be initialised on any number of sites. Examples for four and five sites are shown below on the left and right.\\

\begin{minipage}[t]{0.45\linewidth}
\noindent\textit{Cycle example (\(n=4\)).}
\begin{lstlisting}[style=pyclasses]
motif_1 = Qcycle(stride=2, step=1, mapping=eZY)
tn = Qinit(4) + motif_1
plot_circuit(tn)
\end{lstlisting}
\centering
\includegraphics[width=\linewidth,height=.25\textheight,keepaspectratio]{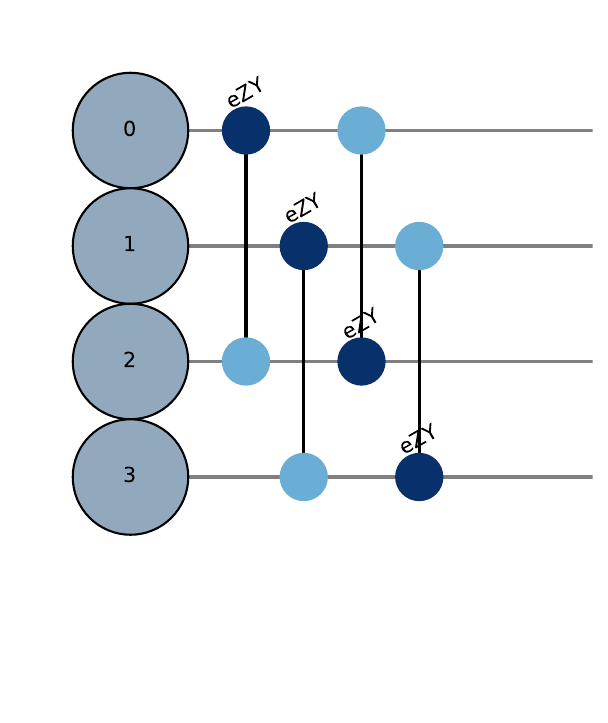}
\vspace{-25pt}
\captionof{figure}{\emph{Motif\_1}, a cycle on \(N=4\) sites.}\label{fig:cycle-n4}
\end{minipage}\hfill
\begin{minipage}[t]{0.45\linewidth}
\noindent\textit{Same motif (\(n=5\)).}
\begin{lstlisting}[style=pyclasses]
tn = Qinit(5) + motif_1
plot_circuit(tn)
\end{lstlisting}
\centering
\includegraphics[width=\linewidth,height=.25\textheight,keepaspectratio]{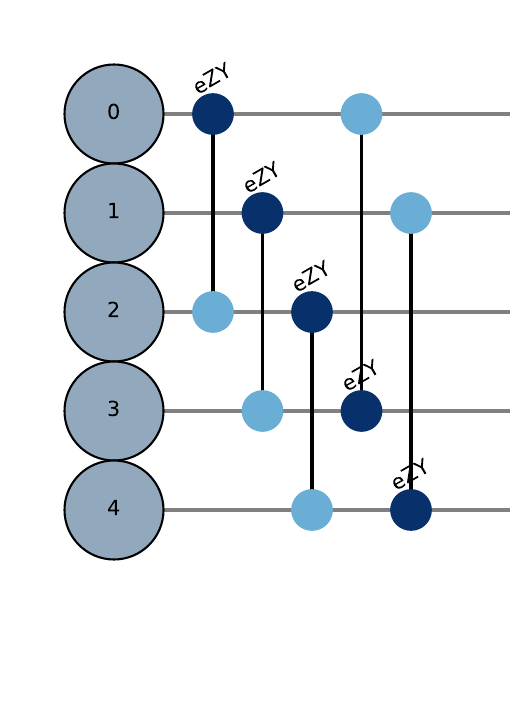}
\vspace{-25pt}
\captionof{figure}{\emph{Motif\_1} on \(N=5\) sites.}\label{fig:cycle-n5}
\end{minipage}\\\\\\
In Fig.~\ref{fig:arity_1-n5} we show an example of a \emph{cycle}, named \emph{motif\_2}, that has an associated tensor of rank two. To its right in Fig.~\ref{fig:pivot} we show an example of a different primitive, a \emph{pivot}. Here, we use the same associated tensor as for \emph{motif\_1}, the pattern is such that the first site is connected to all other sites, see the package documentation \cite{lourensHierarqcalGithub2024} for details.\\\\
\begin{minipage}[t]{0.45\linewidth}
\noindent\textit{Rank 2 tensor (\(n=5\)).}
\begin{lstlisting}[style=pyclasses]
    motif_2 = Qcycle(step=2, mapping=eY)
    tn = Qinit(5) + motif_2
    plot_circuit(tn)
    \end{lstlisting}
    \centering
    \includegraphics[width=\linewidth,height=.25\textheight,keepaspectratio]{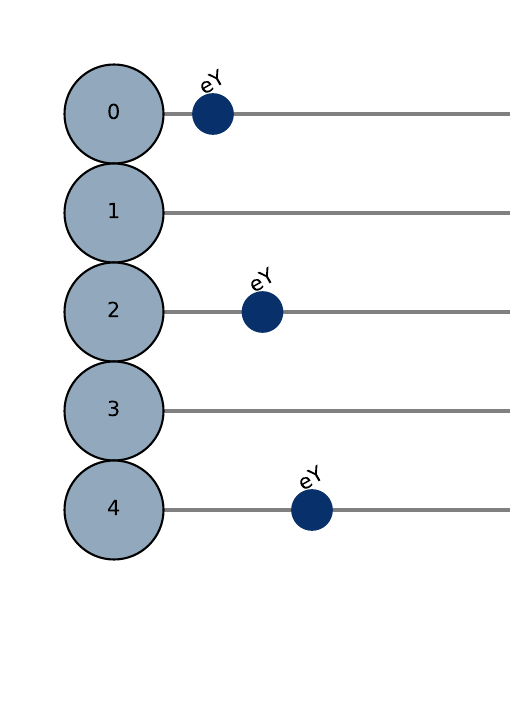}
    \vspace{-25pt}
    \captionof{figure}{\emph{Motif\_2}, a cycle with a rank $2$ tensor.}\label{fig:arity_1-n5}
\end{minipage}\hfill
\begin{minipage}[t]{0.45\linewidth}
\noindent\textit{Pivot (\(n=5\)).}
\begin{lstlisting}[style=pyclasses]
motif_pivot = Qpivot("1*", mapping=eZY)
tn = Qinit(5) + motif_pivot
plot_circuit(tn)
\end{lstlisting}
\centering
\includegraphics[width=\linewidth,height=.25\textheight,keepaspectratio]{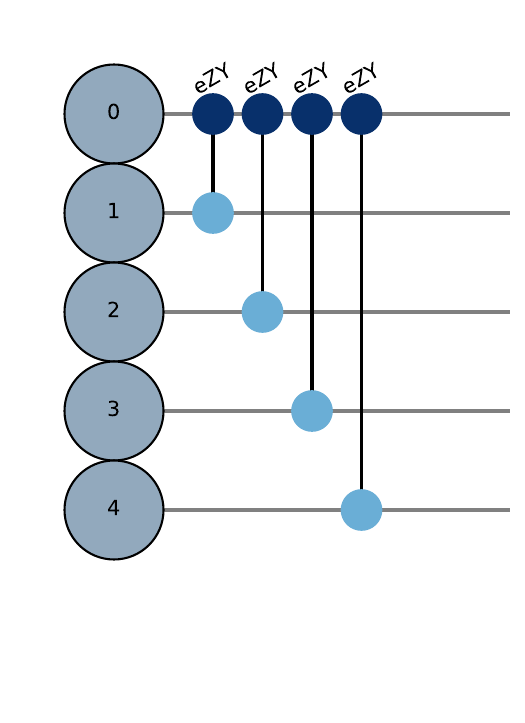}
\vspace{-25pt}
\captionof{figure}{Pivot example.}\label{fig:pivot}
\end{minipage}\\\\\\
As noted in the main text, since a tensor network is itself a tensor, it can again be associated with an edge generation pattern, thereby forming a new primitive, and allowing larger networks to be built hierarchically from sub-networks. Below, we provide an example illustrating this, on the left we create  a tensor network called \emph{pivot\_3}, shown in Fig.~\ref{fig:pivot_3}. To its right, we create a primitive, \emph{cycle\_pivot} which takes \emph{pivot\_3} as its associated tensor. The resulting network seen in Fig.~\ref{fig:cycle_pivot} then consists of the sub-network \emph{pivot\_3}\\\\
\begin{minipage}[t]{0.48\linewidth}
\noindent\textit{Pivot initialised on $3$ sites}
\begin{lstlisting}[style=pyclasses]
    pivot_3 = Qinit(3,name="pivot_3") + motif_pivot
    plot_circuit(pivot_3)
\end{lstlisting}
    \centering
    \includegraphics[width=\linewidth,height=.25\textheight,keepaspectratio]{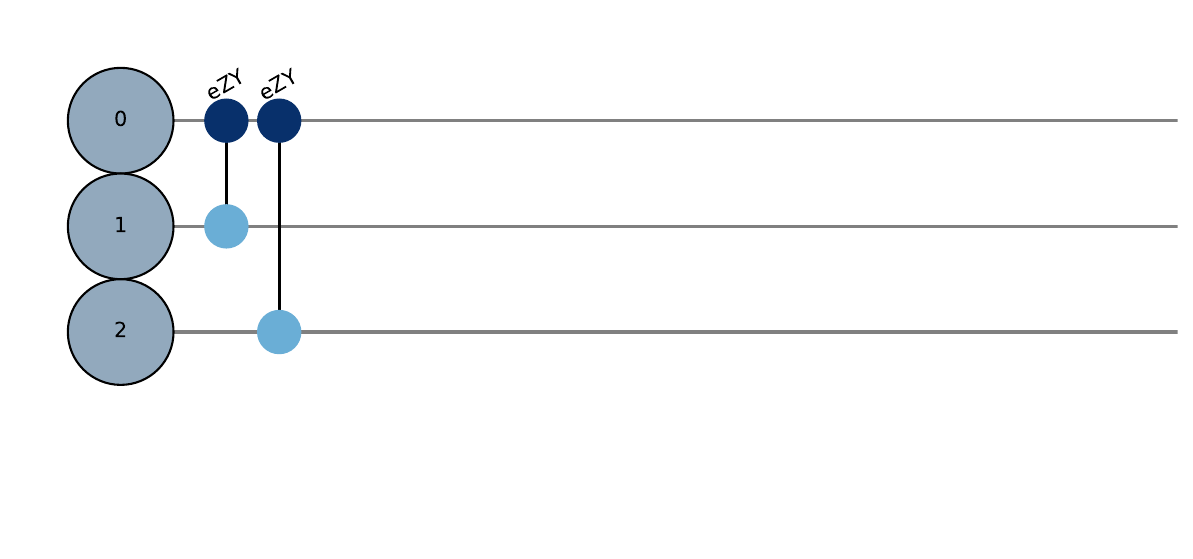}
    \vspace{-25pt}
    \captionof{figure}{The \emph{pivot\_3} tensor network}\label{fig:pivot_3}
\end{minipage}\hfill
\begin{minipage}[t]{0.45\linewidth}
\noindent\textit{Cycle of the pivot}
\begin{lstlisting}[style=pyclasses]
cycle_pivot = Qcycle(step=2, mapping=pivot_3)
tn = Qinit(7) + cycle_pivot
plot_circuit(tn)
\end{lstlisting}
\centering
\includegraphics[width=\linewidth,height=.25\textheight,keepaspectratio]{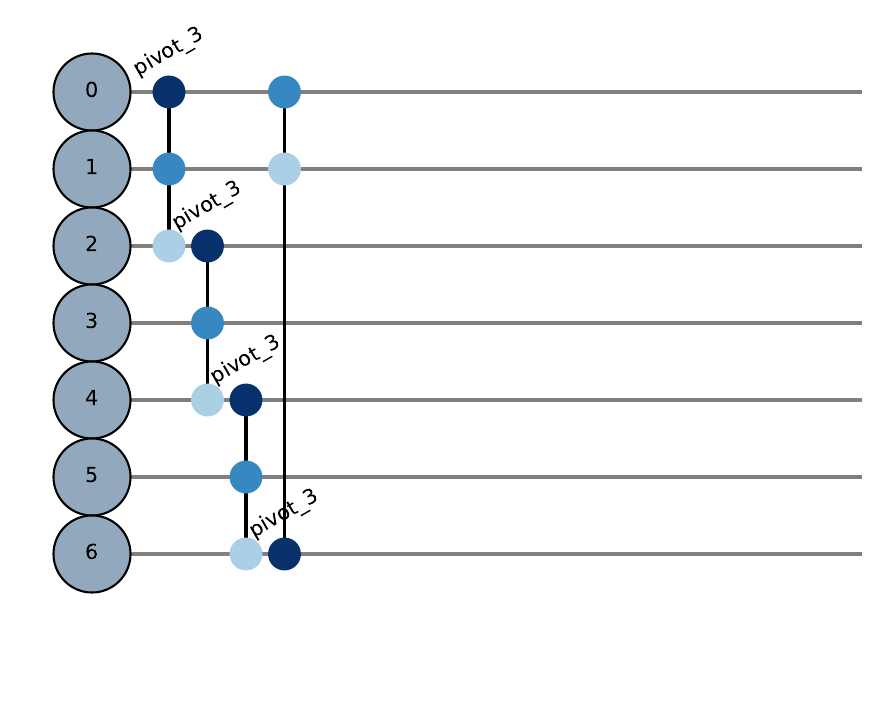}
\vspace{-25pt}
\captionof{figure}{A cycle of \emph{pivot\_3}}\label{fig:cycle_pivot}
\end{minipage}\\\\
In Figs.~\ref{fig:mutation} and~\ref{fig:crossover} we show two simple examples of genetic operations that are used in the search algorithm. During the search, the winning motif of a tournament is mutated, which amounts to a random primitive in the motif having one of its size-independent properties altered. For example, \emph{motif\_1} from Fig.~\ref{fig:cycle-n5} consists of only one primitive and in Fig.~\ref{fig:mutation} its \emph{mapping} property gets altered from $e^{-i\frac{\theta}{4}Z\otimes Y}$ to $e^{-i\frac{\theta}{4}X\otimes Y}$.\\
The two winners in a tournament are also crossed over. Generally we take two random sections from each motif and combine them. In Fig.~\ref{fig:crossover} we show an example of a crossover between \emph{motif\_1} and \emph{motif\_2}. In this example each motif consists of only one primitive, so that the crossover is simply the combination of those two.
\begin{minipage}[t]{0.47\linewidth}
\noindent\textit{Mutation (\(n=5\)).}
\begin{lstlisting}[style=pyclasses]
motif_1_m = Qcycle(stride=2, step=1, mapping=eXY)
tn = Qinit(5) + motif_1_m
plot_circuit(tn)
\end{lstlisting}
\centering
\includegraphics[width=\linewidth,height=.25\textheight,keepaspectratio]{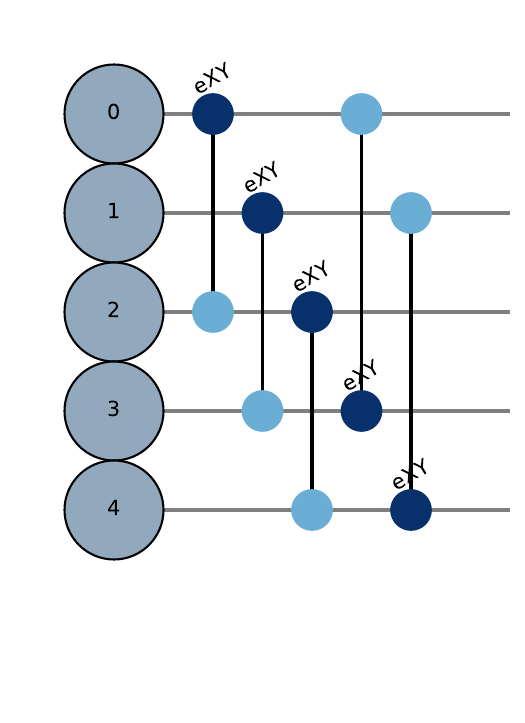}
\vspace{-25pt}
\captionof{figure}{Mutation of \emph{motif\_1}.}\label{fig:mutation}
\end{minipage}\hfill
\begin{minipage}[t]{0.43\linewidth}
\noindent\textit{Crossover (\(n=9\)).}
\begin{lstlisting}[style=pyclasses]
motif_cross = motif_1 + motif_2
tn = Qinit(9) + motif_cross
plot_circuit(tn)
\end{lstlisting}
\centering
\includegraphics[width=\linewidth,height=.25\textheight,keepaspectratio]{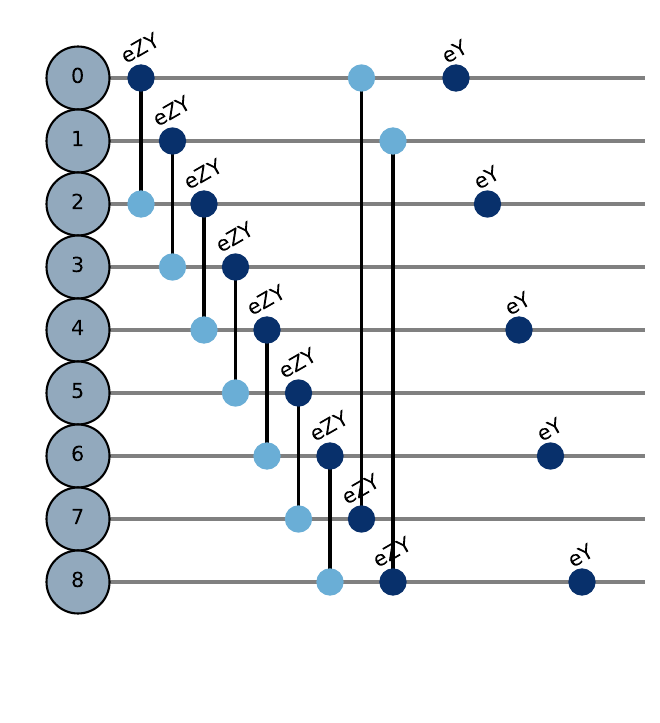}
\vspace{-25pt}
\captionof{figure}{Crossover of \emph{motif\_1} and \emph{motif\_2}.}\label{fig:crossover}
\end{minipage}\\\\
Finally, we show an example of the \emph{mask} primitive, which hides physical sites from subsequent motifs. In Fig.~\ref{fig:mask} we show a mask which always hides the top half of the chain. The \emph{mask} is followed by a \emph{pivot}, and these two are repeated three times. \\
\noindent\textit{Mask (\(n=8\)).}
\begin{lstlisting}[style=pyclasses]
motif_mask = Qmask("!*")
tn = Qinit(8) + (motif_pivot + motif_mask) * 3
plot_circuit(tn)
\end{lstlisting}
\begin{center}
\includegraphics[width=.82\linewidth,height=.35\textheight,keepaspectratio]{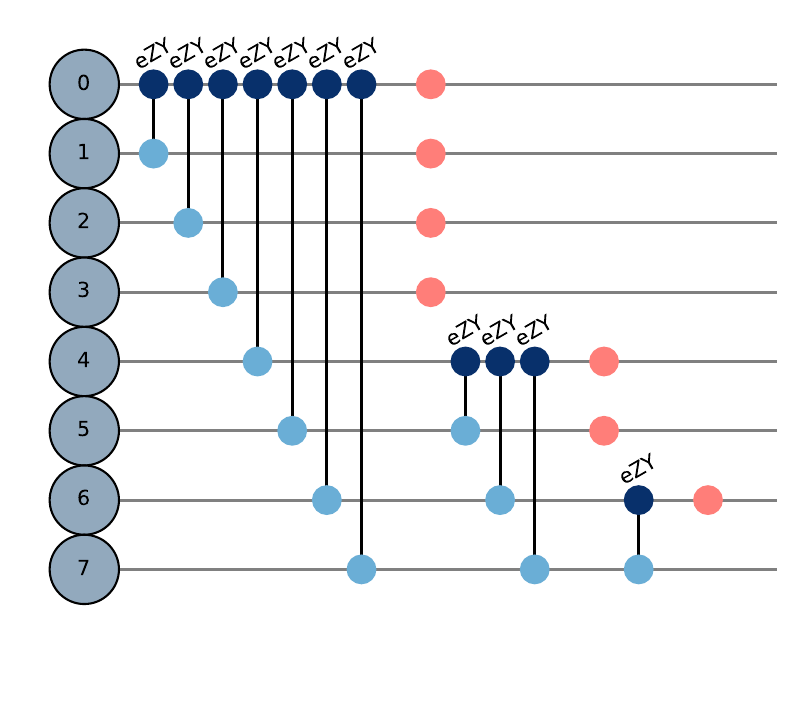}
\captionof{figure}{Example of Mask primitive.}\label{fig:mask}
\end{center}

\section{Growth of bond dimension}
Generally speaking for any MPS, the application of the operator $C^\theta_{ij}$ will increase the bond dimension between sites $i$ and $j$ by a factor of two. This is because $C^\theta_{ij}$ adds a new virtual bond between the sites $i$ and $j$ which we can explicitly capture as a contraction between two two-dimensional vectors of operators,
\begin{align}
    C_{ij}^{\theta} &= \begin{bmatrix} I_i &  Z_i \end{bmatrix}\begin{bmatrix}cI_j \\ isY_j \end{bmatrix} =cI_iI_j + isZ_iY_j. \label{eq:supp_cij_mpo}
\end{align}
Here it is understood that matrix multiplication is performed during the contraction process. Since every site in our ansatz shares one control and one target from all the $C^\theta_{ij}$ operators in Eq.~\eqref{eq:supp_anzcircuit}, every site has two virtual indices each of dimension two. Therefore our state can be written as a matrix product state (MPS) with bond dimension $D=2$. To explicitly see how $C_{ij}$ introduces virtual bonds, consider an arbitrary product state $\ket{\rho}$,
\begin{align}
    \ket{\rho}
       &= \sum_{\vec{\sigma}}
          \Bigl( \prod_{k=0}^{N-1} p_k(\sigma_k) \Bigr)
          \, \ket{\vec{\sigma}}
    \end{align}
where $p_k(\sigma_k)\in \mathbb{C}$ are order-0 tensors. Applying $C_{ij}^{\theta}$ on $\ket{\rho}$ turns the scalars $p_i$ and $p_j$ into the vectors $\mathbf{p_i}$ and $\mathbf{p_j}$ as follows:
\begin{align}
    C_{ij}^{\theta}\ket{\rho}
    &= \sum_{\vec{\sigma}}
       \Bigl( \prod_{k=0}^{N-1} p_k(\sigma_k) \Bigr)\!
       \Bigl(
           \begin{bmatrix} I_i & Z_i \end{bmatrix}
           \begin{bmatrix} c I_j \\ is\,Y_j \end{bmatrix}
       \Bigr)
       \ket{\sigma_i\sigma_j}\!
       \otimes \ket{\vec{\sigma}^{\prime}},\\
    &= \sum_{\vec{\sigma}}
        \Bigl( \prod_{k=0}^{N-1} p_k(\sigma_k) \Bigr)\!
        \Bigl(
            \begin{bmatrix} \ket{\sigma_i} & \ket{-\sigma_i} \end{bmatrix}
            \begin{bmatrix} c\,\ket{\sigma_j} \\ is\,\sigma_j\ket{\sigma_j} \end{bmatrix}
        \Bigr)
        \otimes \ket{\vec{\sigma}^{\prime}},\\
    &= \sum_{\vec{\sigma}}
        \Bigl( \prod_{k\neq i,j} p_k(\sigma_k) \Bigr)\!
        \Bigl(
            \begin{bmatrix} p_i(\sigma_i)\ket{\sigma_i} & p_i(\sigma_i)\ket{-\sigma_i} \end{bmatrix}
            \begin{bmatrix} c\,p_j(\sigma_j)\ket{\sigma_j} \\ is\,\sigma_jp_j(\sigma_j)\ket{\sigma_j} \end{bmatrix}
        \Bigr)
        \otimes \ket{\vec{\sigma}^{\prime}},\\
    &= \sum_{\vec{\sigma}}
        \Bigl( \prod_{k\neq i,j} p_k(\sigma_k) \Bigr)
        \Bigl (
            \begin{bmatrix} p_i(\sigma_i) & p_i(-\sigma_i) \end{bmatrix}
            \begin{bmatrix} c\,p_j(\sigma_j) \\ is\,\sigma_j p_j(\sigma_j) \end{bmatrix}
        \Bigr )
        \ket{\vec{\sigma}},\label{eq:supp_regroup}\\
    &\equiv \sum_{\vec{\sigma}}
    \Bigl( \prod_{k\neq i,j} p_k(\sigma_k) \Bigr)
    \mathbf{p}_i(\sigma_i)^\dagger\mathbf{p}_j(\sigma_j)
    \ket{\vec{\sigma}}.
\end{align}
Here we obtained Eq.~\eqref{eq:supp_regroup} by regrouping terms that share the same $\sigma_i$ in the total sum. Similarly, applying one $C_{ij}^{\theta}$ after the other leads to
\begin{align}
    C_{jl}^{\theta}C_{ij}^{\theta}\ket{\rho}
    &= \sum_{\vec{\sigma}}
    \Bigl( \prod_{k\neq i,j} p_k(\sigma_k) \Bigr)
    \mathbf{p}_i(\sigma_i)^\dagger\mathbf{p}_j(\sigma_j)\!
    \Bigl(
           \begin{bmatrix} I_j & Z_j \end{bmatrix}
           \begin{bmatrix} c I_l \\ is\,Y_l \end{bmatrix}
       \Bigr)
       \ket{\sigma_i\sigma_l}\!
       \otimes \ket{\vec{\sigma}^{\prime}},\\
    &= \sum_{\vec{\sigma}}
    \Bigl( \prod_{k\neq i,j} p_k(\sigma_k) \Bigr)
    \mathbf{p}_i(\sigma_i)^\dagger\!
    \Bigl (
    \begin{bmatrix}  \mathbf{p}_j(\sigma_j) & 0 \end{bmatrix}\otimes\ket{\sigma_j}         
    + \begin{bmatrix} 0 &  \mathbf{p}_j(\sigma_j) \end{bmatrix}\otimes \ket{-\sigma_j}
    \Bigr ) \begin{bmatrix} c\,p_l(\sigma_l) \\ is\,\sigma_l p_l(\sigma_l) \end{bmatrix}
    \otimes \ket{\sigma_l,\vec{\sigma}^{\prime}},\\
    &= \sum_{\vec{\sigma}}
       \Bigl( \prod_{k\neq i,j,l} p_k(\sigma_k) \Bigr)\!
       \mathbf{p}_i(\sigma_i)^\dagger\!
       \Bigl (
           \begin{bmatrix} \mathbf{p}_j(\sigma_j) & \mathbf{p}_j(-\sigma_j) \end{bmatrix}
           \begin{bmatrix} c\, p_l(\sigma_l) \\ is\,\sigma_l  p_l(\sigma_l) \end{bmatrix}
       \Bigr )
       \ket{\vec{\sigma}},\\
    &\equiv \sum_{\vec{\sigma}}
    \Bigl( \prod_{k\neq i,j,l} p_k(\sigma_k) \Bigr)\!
    \mathbf{p}_i(\sigma_i)^\dagger\!
    P_j(\sigma_j)
    \mathbf{p}_l(\sigma_l)
    \ket{\vec{\sigma}}. \label{eq:matrix_from_cijcjl}
\end{align}
Here $C^\theta_{ij}$ introduced a virtual bond between sites $i$ and $j$ and $C^\theta_{jl}$ introduced a virtual bond between sites $j$ and $l$. Resulting in site $j$ having two virtual indices, each of dimension two.
\section{Alternative calculation of the MPS}
The quickest way to obtain an MPS for ansatzes of the form in Eq.~\eqref{eq:supp_anzcircuit} is by considering each site on its own, explicitly capturing the virtual bonds introduced by operators and contracting along the physical index. For illustration purposes, we show how to obtain the $A^\pm_z$ matrices from Eq.~\eqref{eq:ABpm_matrices}, which represent the MPS matrices for the $Z$ basis version of the ansatz. Using the decomposition from Eq.~\eqref{eq:supp_cij_mpo} of the $C^\theta_{ij}$ operator, we obtain the effective tensor describing any site $i\neq0$ as
\begin{align}
 A_z&=C_{i,i+1}C_{i-1,i}R^{\theta+\phi}_i\ket{z,+}\\
 &=  \begin{bmatrix} I_i &  Z_i \end{bmatrix} \otimes \begin{bmatrix} cI_i \\ isY_i \end{bmatrix}  \bigl( e^{-i\frac{\theta+\phi}{2}Y_i}\bigr) \ket{z,+},\\
 &=  \begin{bmatrix} cI_i  &  cZ_i\\ isY_i & sX_i\end{bmatrix} \Bigl(d\ket{z,+}+t\ket{z,-}\Bigr),\\
 &=  \begin{bmatrix} c\Bigl(d\ket{z,+}+t\ket{z,-}\Bigr) &  c\Bigl(d\ket{z,+}-t\ket{z,-}\Bigr)\\ is\Bigl(id\ket{z,-}+it\ket{z,+}\Bigr)& s\Bigl(d\ket{z,-}+t\ket{z,+}\Bigr)\end{bmatrix},\\
 &=  \begin{bmatrix} cd &  cd\\ st& st\end{bmatrix}\ket{z,+}+\begin{bmatrix} ct &  -ct\\ -sd& sd\end{bmatrix}\ket{z,-}.\label{eq:Az_pm}
\end{align}
Here it is understood that virtual indices which connects to sites $i-1$ and $i+1$ are written explicitly as vectors or matrices and a contraction is performed along the physical index for site $i$. Similarly, for site $i=0$ we have
\begin{align}
    B_z &= C^\theta_{N-1,0}R^{\theta}_0C^\theta_{0,1}R^{\phi}_0\ket{z,+}\\
    &=  \begin{bmatrix} cI_0 \\ isY_0 \end{bmatrix} \otimes \bigl( e^{-i\frac{\theta}{2}Y_0}\bigr)\begin{bmatrix} I_0 &  Z_0 \end{bmatrix}  \bigl( e^{-i\frac{\phi}{2}Y_0}\bigr) \ket{z,+},\\
    &=\begin{bmatrix} cI_0 \\ isY_0 \end{bmatrix} \otimes \begin{bmatrix}  e^{-i\frac{\theta}{2}Y_0}  &  Z_0 e^{+i\frac{\theta}{2}Y_0}  \end{bmatrix}  \bigl( e^{-i\frac{\phi}{2}Y_0}\bigr)\ket{z,+},\\
    &=\begin{bmatrix} ce^{-i\frac{\theta+\phi}{2}Y_0}  &  cZ_0 e^{+i\frac{\theta-\phi}{2}Y_0}, \\ \
         isY_0 e^{-i\frac{\theta+\phi}{2}Y_0}  &  -sX_0e^{+i\frac{\theta-\phi}{2}Y_0} \end{bmatrix} \ket{z,+}\\
    &=\begin{bmatrix} c\Bigl(d\ket{z,+}+t\ket{z,-}\Bigr)  &  cZ_0 \Bigl(e\ket{z,+}-u\ket{z,-}\Bigr) \\ \
    isY_0 \Bigl(d\ket{z,+}+t\ket{z,-}\Bigr) &  -sX_0\Bigl(e\ket{z,+}-u\ket{z,-}\Bigr) \end{bmatrix},\\
    &=\begin{bmatrix} c\Bigl(d\ket{z,+}+t\ket{z,-}\Bigr)  &  c\Bigl(e\ket{z,+}+u\ket{z,-}\Bigr) \\ \
        is\Bigl(id\ket{z,-}+-it\ket{z,+}\Bigr) &  -s\Bigl(e\ket{z,-}-u\ket{z,+}\Bigr) \end{bmatrix},\\
    & = \begin{bmatrix} cd  &  ce \\  st & su \end{bmatrix}\ket{z,+}+ \begin{bmatrix} ct  &  cu \\  -sd & -se \end{bmatrix} \ket{z,-}.\label{eq:Bz_pm}
\end{align}
From Eqs.~\eqref{eq:Az_pm} and ~\eqref{eq:Bz_pm} we recognize the $A^\pm_z,B^\pm_z$ matrices from Eq.~\eqref{eq:ABpm_matrices}.
\end{document}